\documentclass[a4paper,aps,noeprint,twocolumn,secnumarabic,
  balancelastpage,amsmath,amssymb,nofootinbib,longbibliography,
  preprint,10pt
]{revtex4-1}
\usepackage{pifont}
\usepackage{graphicx}
\usepackage{dcolumn}
\usepackage{subfigure}
\usepackage{booktabs}
\usepackage[normalem]{ulem}
\usepackage{siunitx}
\usepackage{bm}
\usepackage[colorlinks=true,breaklinks=true,allcolors=blue]{hyperref}
\usepackage{url}
\usepackage{xcolor}
\usepackage{enumitem}
 
\makeatletter

\newcommand{\me}[1]{\left\langle #1 \right\rangle }

\makeatletter
\let\cat@comma@active\@empty
\makeatother

\AtBeginDocument{%
    \newwrite\bibnotes
    \def\bibnotesext{Notes.bib}
    \immediate\openout\bibnotes=\jobname\bibnotesext
    \immediate\write\bibnotes{@CONTROL{REVTEX41Control}}
    \immediate\write\bibnotes{@CONTROL{%
    apsrev41Control,author="08",editor="1",pages="1",title="0",year="1"}}
     \if@filesw
     \immediate\write\@auxout{\string\citation{apsrev41Control}}%
    \fi
}%

\begin{document}

\title{
\texorpdfstring{$BKT$}{BKT} transitions 
in classical and quantum  
long-range systems
}

\author{Guido Giachetti}
\email{ggiachet@sissa.it}
\affiliation{SISSA and INFN Sezione di Trieste, Via Bonomea 265, I-34136 Trieste, Italy}

\author{Andrea Trombettoni}
\affiliation{Department of Physics, University of Trieste, Strada Costiera 11, I-34151 Trieste, Italy}
\affiliation{SISSA and INFN Sezione di Trieste, Via Bonomea 265, I-34136 Trieste, Italy}

\author{Stefano Ruffo}
\affiliation{SISSA and INFN Sezione di Trieste, Via Bonomea 265, I-34136 Trieste, Italy}
\affiliation{Istituto dei Sistemi Complessi, Consiglio Nazionale delle Ricerche, Via Madonna del Piano 10, I-50019 Sesto Fiorentino, Italy}

\author{Nicol\`o Defenu}
\affiliation{Institut f\"ur Theoretische Physik, ETH Z\"urich, Wolfgang-Pauli-Str.\,27 Z\"urich, Switzerland}

\begin{abstract}
\noindent
In the past decades considerable efforts have been made in 
order to understand the critical features of both 
classical and quantum long-range interacting models. The case of the 
Berezinskii-Kosterlitz-Thouless (BKT) universality class, as in the $2d$ classical $XY$ model, is considerably complicated by the presence, for short-range interactions, of a line of renormalization group fixed points. In this paper we discuss a field theoretical treatment of the $2d$ $XY$ model with long-range couplings and we compare it with results from the self-consistent harmonic approximation. 
These methods lead to a rich phase diagram, where both power-law BKT scaling and spontaneous symmetry breaking appear for the same (intermediate) decay rates of long-range interactions. \textcolor{black}{We also discuss the Villain approximation for the $2d$ $XY$ model with power-law couplings, providing hints 
that, in the long-range regime, it 
fails to reproduce the correct 
critical behavior.}
The 
\textcolor{black}{obtained results are then} 
applied to the long-range quantum $XXZ$ spin chain at zero temperature. We discuss the relation between the phase diagrams of the two models and we give predictions about the scaling of the order parameter of the quantum chain close to the transition.
\end{abstract}

\maketitle

\section{Introduction}
\noindent
In the context of statistical mechanics and condensed matter, it is well known that the presence of long-range (LR) interactions, such as slow-decaying couplings with power-law behavior at large distances, gives rise to plenty of new physical phenomena in the equilibrium and non-equilibrium properties of both classical\,\cite{campa2014physics} and quantum\,\cite{defenu2021longrange} systems. These properties have recently sparkled
a new wave of interest, due \textcolor{black}{to} the possibility of experimental realizations in atomic, molecular
and optical (AMO) systems\,\cite{haffner2008quantum,lahaye2009physics,saffman2010quantum, ritsch2013cold,bernien2017quantum,monroe2021programmable,defenu2021longrange,mivehvar2021cavity}. 

If the decay of the interaction is slow enough, LR effects influence the universal critical properties and even induce a 
spontaneous symmetry breaking (SSB) in a low-dimensional system, since the celebrated Hohenberg-Mermin-Wagner theorem does not hold in presence of LR couplings, as discussed since the 
\textcolor{black}{classic} papers\,\cite{Dyson1969,Thouless1969,Kosterlitz76}. 

A qualitative and quantitative understanding
of the critical properties of LR interacting systems follows from the Sak criterion\,\cite{sak1973recursion}. Let us consider power-law couplings of the form $\sim 1/r^{d+\sigma}$, where $d$ is the system dimension. For the sake of clarity, in the following considerations we explicitly refer to classical $O(n)$ models\,\cite{defenu2015fixed}. In the gaussian non-interacting case ($n \to \infty$), for large distances and low energies\textcolor{black}{,} one has to compare the short-range (SR) critical propagator, $1/k^2$, with the LR one,
$1/k^\sigma$. However, interacting systems possess a finite anomalous
dimension $\eta$ at the critical point, so that the  propagator scales as $1/k^{2-\eta}$ \cite{cardybook}.
Denoting by $\eta_{\rm sr}$ the anomalous dimension in the SR limit
of the theory, i.e. $\sigma\to\infty$, one concludes that {\it at criticality}
for $\sigma>2-\eta_{\rm sr}$ the system behaves
as its SR counterpart, while for $\sigma<2-\eta_{\rm sr}$ the LR nature of the
couplings modifies the critical behaviour. This \textcolor{black}{argument} is at the heart of 
Sak\textcolor{black}{'s} seminal paper\,\cite{sak1973recursion} and it implies that one can define a
value $\sigma_*$ such that
for $\sigma > \sigma_*$ the system 
is in the universality class of its SR counterpart, 
while for $\sigma < \sigma_*$ the system changes universality class. The value of $\sigma_*$ is given by
\begin{equation}
    \sigma_*=2-\eta_{\rm sr}.
\end{equation}

\textcolor{black}{Sak's} criterion was formulated for classical $O(n)$ models, such as the
Ising model, a playground
where it has been \textcolor{black}{the} subject of \textcolor{black}{thorough} investigation.
Regarding its validity
it is fair to conclude that, \textcolor{black}{to this} date, it has to be considered valid and well
tested, see reviews \cite{Luijten,Defenu2020review} and \textcolor{black}{Refs} therein.
Despite the fact that
there is no rigorous proof for it,  
\textcolor{black}{and} numerical results can only
put an upper bound on its violation, its usefulness is unquestioned. For the
quantum $O(n)$ models, one finds $\sigma_*=2$ \textcolor{black}{at mean-field level} \cite{dutta2001phase}
and one can obtain beyond\textcolor{black}{-}mean-field results by applying renormalization
group techniques \cite{defenu2017criticality}.

\textcolor{black}{Sak's} criterion applies
to all $d$ and $n$, except for $d=2$ and $n=2$, i.e. the $2d$ XY universality class. There, it is well known that 
SSB occurs \textcolor{black}{at low temperature} for $\sigma<2$ \cite{kunz1976first} and, therefore, several questions arise concerning the \textcolor{black}{applicability} of \textcolor{black}{Sak's} criterion in the SR limit, where the model does not present SSB. Moreover, \textcolor{black}{in the $2d$ XY model,} $\eta_{\rm sr}$ -- unlike the other cases covered by \textcolor{black}{Sak's} criterion -- \textcolor{black}{actually depends} 
on \textcolor{black}{the} temperature, \textcolor{black}{raising} further concerns about the 
\textcolor{black}{ applicability} of the criterion \textcolor{black}{to the $2d$ XY model with LR couplings.}

In the recent paper\,\cite{Giachetti2021}, the fate of the BKT line of critical points has been discussed. For $\sigma>2$ simple arguments (not directly applicable to $\sigma<2$) show the existence of the BKT transition\,\cite{Giachetti2021EPL}, while what happens for $\sigma<2$ is considerably more subtle. In Ref.\,\cite{Giachetti2021} SSB was found to persist in the range $7/4<\sigma<2$ up to a finite critical temperature $T_c$, beyond which a quasi-ordered BKT-like phase appears \textcolor{black}{below} a larger temperature $T_{\rm BKT}$. These results are compatible with the application of \textcolor{black}{Sak's} criterion with the temperature dependent $\eta_{\rm sr}$ of the $XY$ model in the SR limit. 

Since the classical $2d$ $XY$ model at finite temperature lies in the same universality class of the quantum $XXZ$ chain at $T=0$ for nearest-neighbors-interactions, a natural question is how to apply these results for the $2d$ LR XY model to the $1d$ LR $XXZ$ chain. We remind that\textcolor{black}{,} in the SR limit, \textcolor{black}{a} mapping \textcolor{black}{can be} obtained via the transfer matrix technique \cite{Mattis1984}. However, this method does not straightforwardly apply to the LR case, nor to any non-local couplings, in general. While it is natural to think the two models to be in the same universality class for fast-decaying couplings as well, it is not clear what happens in the LR regime. We observe that the $2D$ boson gas at finite temperature with $1/r^3$ \textit{density-density} interactions is equivalent to a quantum $XXZ$ chain with $z-z$ LR couplings and it has been found to follow the usual BKT scenario \cite{Filinov2010} with no signatures of SSB.

The LR $1d$ $XXZ$ chain has been studied in Ref.\,\cite{maghrebi2017continuous}, where \textcolor{black}{evidence} were provided showing the existence of three phases in the $T=0$ phase diagram as a function of $\sigma$ and $\Delta$ (with $\Delta$ the anisotropy parameter). One phase features SSB, another is disordered and the third is gapless. The comparison between the  phase diagrams of Ref.\,\cite{Giachetti2021} for the $2d$ classical finite temperature XY model and of Ref.\,\cite{maghrebi2017continuous} for the $1d$ quantum zero temperature $XXZ$ model is one of the motivations of the present paper.

The study of the $2d$ XY model with LR interactions is of great interest \textit{per se}: the model is indeed \textcolor{black}{paradigmatic} for the study of $O(2)$-symmetric systems in $d=2$ 
(\textcolor{black}{including} thin $^{4}$He films\,\cite{Bishop1978}, quasi-2D layered superconductors\,\cite{corson_vanishing_1999, Rendeira_NatPhys07, bilbro_temporal_2011, Lemberger_PRB85_2012, Baity_LB_PRB93_2016},
exciton-polariton systems\,\cite{Nitsche2014}, cold atoms in $2D$ traps\,\cite{dalibard_2006,Murthy_PRL115_2015} and $2D$ electron gases at the interface between insulating oxides in artificial heterostructures\,\cite{reyren_superconducting_2007, Daptary_PhysRevB.94.085104, Monteiro_PhysRevB.96.020504}).

In the present paper, the scenario presented in Ref.\,\cite{Giachetti2021} for the $2d$ $XY$ model is analyzed 
\textcolor{black}{in detail} and 
\textcolor{black}{investigated 
} using different methods: the low-temperature expansion, an extension of the traditional self-consistent harmonic approximation (SCHA) \,\cite{POKROVSKY1973,Pires1997,Giachetti2021EPL} to $\sigma<2$ and renormalization group (RG) field theoretical arguments. The difficulties inherent in the use of the Villain approximation for $\sigma<2$ are pointed out and contrasted with the SR case, where, at variance, the $2d$ Villain and XY model are related \textcolor{black}{through} controllable approximations and lie in the same universality class. Using these results, we closely examine the link between the classical XY model in $2d$ and the quantum $XXZ$ chain, leading to a 
\textcolor{black}{detailed} analysis of the quantum model. The comparison with the results of Ref.\,\cite{maghrebi2017continuous} is presented. \textcolor{black}{Comments on the BKT transition in LR spin-$S$ quantum chains are also presented.}

The paper is structured as follows: after presenting the model in Sec. II, in Sec. III we describe its low-temperature limit. In Sec. IV and V, we examine the peculiar nature of the crossover between LR and SR regimes. In particular, Sec. IV is devoted to extending the self-consistent approximation to the proper $\sigma<2$ LR regime; while in Sec V we examine the field-theory picture. 
In Sec. VI we discuss the relation between our analysis and the Villain method used to understand the BKT phase transition in the SR regime, showing the difficulties encountered to describe this crossover. Finally, in Sec. VII we examine the relationship between the $d=2$ classical $XY$ model and the $1d$ quantum $XXZ$ chain, showing how our results can be used to make predictions about the $T=0$ phase diagram of the latter. 

\section{The \texorpdfstring{$XY$}{xy}
model}
The model consists of a set of $N$ planar rotators, arranged in a  $2$-dimensional square lattice, interacting through the Hamiltonian
\begin{equation}\label{model_xy}
\beta H = \frac{1}{2} \sum_{\mathbf{i},\mathbf{j}} J(r) \left[1 - \cos ( \theta_{\mathbf{i}} - \theta_{\mathbf{j}})\right]
\end{equation}
with $\mathbf{i}, \mathbf{j} \in \mathbb{Z}^2$, $\mathbf{r} = \mathbf{i}-\mathbf{j}$, $r = |\mathbf{r}|$, 
\textcolor{black}{$J(r) \sim J {r^{-2-\sigma}}$} for $r\gg 1$ where $J = \beta J_0$. \textcolor{black}{Here we used a particular instance of the general convention $J(r) \sim r^{- d  - \sigma}$ used in  the $d$-dimensional case (see e.g. \cite{defenu2017criticality}), which assures that the thermodynamical quantities remain additive for any $\sigma > 0$ \cite{campa2014physics}.} The interaction between the spins 
$\textcolor{black}{\mathbf{S}_{\mathbf{x}}} \equiv \left(\cos \theta (\mathbf{x}), \sin \theta (\mathbf{x}) \right)$ is invariant under global $O(2)$ rotations. In what follows we take our energy units so that $J_0=1$
\textcolor{black}{ and we will limit ourself to the $\sigma > 0$ case}.

In the SR regime, although the possibility of a SSB is ruled out by the \textcolor{black}{Hohenberg}-Mermin-Wagner theorem, the low-temperature phase is characterized by quasi-long-range-order, i.e. power-law behavior in the connected correlations functions with a temperature-dependent exponent. The transition between this phase and the disordered, high-temperature one, i.e. the well-known BKT mechanism, may be understood in terms of \textcolor{black}{the unbinding of} topological defects, 
leading to the celebrated mapping to a Coulomb gas of charged particles\,\cite{kosterlitz2017nobel,berezinsky,KT,K,review}.

When the decay of the couplings is slow enough, we expect a SSB phase to appear at low temperature. The accepted criterion for understanding whether the LR interactions do actually modify the critical behavior is the one provided by Sak \cite{sak1973recursion}. However, as discussed in the introduction, this criterion cannot be straightforwardly applied to the BKT mechanism, due to the temperature-dependent anomalous dimension $\eta_{\rm sr}$.
Moreover, also the duality construction\,\cite{savit}, which allows the mapping between the $2d$ SR XY model, the Coulomb gas\,\cite{Minnhagen, Gulacsi1998} and the sine-Gordon model\,\cite{Amit, Gulacsi1998}, breaks down, already in the case of next-to-nearest-neighbors couplings. The other way of obtaining the same mappings is the so-called Villain approximation\,\cite{Villain1974,Villain1975,Jose1977} (i.e. the substitution of Hamiltonian \eqref{model} with a quadratic one which takes into account the phase periodicity). However, as shown in Sec. VI of this paper, there is reason to believe that this latter Hamiltonian is no longer in the same universality class as soon as the interaction becomes LR, so that the Coulomb gas picture 
\textcolor{black}{cannot be straightforwardly used to describe} 
the physics of the $2d$ LR XY model. 

\section{Low temperature expansion}
\noindent
In analogy with the SR regime, we expect the low-temperature behavior of the model to be well described by an approximation \emph{a lá} Berezinskii\,\cite{berezinsky}. Since we expect our thermal fluctuations to be small in this regime, we expand the cosine in the Hamiltonian \eqref{model} to second order, obtaining, up to an immaterial constant term,
\begin{equation}\label{model}
\beta H \sim \frac{1}{4} \sum_{\mathbf{i},\mathbf{j}} J(r) ( \theta_{\mathbf{i}} - \theta_{\mathbf{j}})^2.
\end{equation}
Being this theory quadratic and translationally invariant, we can 
\textcolor{black}{diagonalize} 
the above Hamiltonian by means of a Fourier transform
\begin{equation}
\theta_{\mathbf{q}} = \frac{1}{\sqrt{N}} \sum_{\mathbf{j}} e^{ - i \mathbf{q} \cdot \mathbf{j}} \ \theta_{\mathbf{j}} \hspace{1cm} \theta_{\mathbf{j}} = \frac{1}{\sqrt{N}} \sum_{\mathbf{q} \in IBZ} e^{  i \mathbf{q} \cdot \mathbf{j}} \ \theta_{\mathbf{q}}.
\end{equation}
We obtain
\begin{equation} \label{diagHB}
\beta H \sim \frac{1}{2} \sum_{\mathbf{q} \in IBZ} K(\mathbf{q}) |\theta_{\mathbf{q}}|^2,
\end{equation}
where we introduced 
\begin{equation} \label{Kq}
K(\mathbf{q}) = \sum_{\mathbf{r}} J (\mathbf{r}) \Bigl( 1 - \cos(\mathbf{q} \cdot \mathbf{r}) \Bigr).
\end{equation}
If we are interested in the long-wavelength modes, we can approximate the above sum with a continuous integral, obtaining: 
\begin{equation}
\label{Kqc}
    K(q) = \int_{r>a} {d^2  \mathbf{r}  \textcolor{black}{J(r)}  \Bigl(1- \cos(\mathbf{q} \cdot \mathbf{r}) \Bigr)}.
\end{equation}
As shown in Appendix \ref{AppA}, this quantity grows as $K(q) \sim q^2$ if $\sigma> 2$, while it shows a non-analitic behavior $K(q) \sim q^{\sigma}$ for $\sigma <2$. The difference between the two regimes can be fully appreciated if we consider the two-point correlation function:
\begin{equation}
    \me{\mathbf{s}_{\mathbf{r}} \cdot \mathbf{s}_{0}} = \me{\cos(\theta_{\mathbf{r}} - \theta_0)}  . 
\end{equation}
Being the theory quadratic in 
\textcolor{black}{this}
approximation, \textcolor{black}{the above quantity} can be \textcolor{black}{evaluated} quite easily 
\textcolor{black}{ by exploiting} the identity 
\textcolor{black}{$\me{\cos A}_0 = e^{- \frac{1}{2} \me{A^2}_0}$}. From Eq.\,\eqref{diagHB}, \textcolor{black}{it} follows immediately that $\me{\theta_\mathbf{q}\theta_\mathbf{q'}}_0 = \delta_{\mathbf{q} + \mathbf{q'}} K(\mathbf{q})^{-1}$ so that we find $\me{\cos(\theta_{\mathbf{r}} - \theta_0)} = e^{-G(\mathbf{r})}$ with
\begin{equation} \label{Gr}
G(\mathbf{r}) = \frac{1}{2} \me{( \theta_{0} - \theta_{\mathbf{r}})^2}_0 = \frac{1}{N} \sum_{\mathbf{q} \in IBZ} \frac{1 - \cos(\mathbf{q} \cdot \mathbf{r})}{K(\mathbf{q})}.
\end{equation} 
The long-distance behavior can be once again captured by replacing the sum with an integral
\begin{equation} \label{Grc}
    G(r) = a^2 \int_{q < \Lambda} \frac{d^2 \mathbf{q}}{(2 \pi)^2} \frac{1- \cos(\mathbf{q} \cdot \mathbf{r})}{K(q)},
\end{equation}
where we approximated the first Brillouin zone with a sphere of radius $\Lambda = \sqrt{8 \pi} a^{-1}$, so that its volume is preserved, i.e. $\int_{q < \Lambda} d^2 \mathbf{q} = (2 \pi)^2 a^{-2}$. As shown in Appendix \ref{AppB}, the asymptotic behavior of $G(r)$ depends on $\sigma$. In particular, if $\sigma>2$ we have
\begin{equation}
    G(r) \sim \eta(J) \ln \frac{r}{a} + A J^{-1} 
\end{equation}
with $\eta(J) = p/J$, and $A,p$ are non-universal constants (see a study of these constants in \cite{Giachetti2021EPL}). If $\sigma < 2$ instead, we have that
\begin{equation}
    G(r) \sim A J^{-1}.
\end{equation}
Then, depending on whether $\sigma <2$ or $\sigma >2$ we have that the correlation $\me{\mathbf{s}_{\mathbf{r}} \cdot \mathbf{s}_{0}} \sim \rm const$ or $\me{\mathbf{s}_{\mathbf{r}} \cdot \mathbf{s}_{0}} \sim r^{- \eta(J)}$ respectively. In the latter case then, we recover the SR low-temperature BKT behavior\,\cite{berezinsky}, in which the correlations decay as a power law with a temperature-dependent exponent. The former case, instead, give\textcolor{black}{s} rise to 
\textcolor{black}{a finite-magnetization ordered phase, with}
\begin{equation} \label{mBer}
    m^2 = \lim_{r \rightarrow \infty} \me{\mathbf{s}_{\mathbf{r}} \cdot \mathbf{s}_{0}} = \lim_{r \rightarrow \infty} e^{-G(r)}. 
\end{equation}

This argument then suggests that $\sigma_{*} = 2$. 
If compared with Sak's criterion, this implies an effective anomalous exponent $\eta = 0$. Let us notice, however, how \textcolor{black}{the} low-temperature approximation \textit{per se} is unable to describe the topological configurations, since $\theta_{\mathbf{i}}$ is no longer a phase, and thus is no longer defined up to multiples of $2 \pi$. As a consequence, even in the SR case, the approximation is unable to \textcolor{black}{correctly 
reproduce all the phenomenology} of the BKT transition.

\section{Self-consistent approach}
\label{sec_scha}
For the nearest-neighbor case, the results of the low-temperature expansion 
\textcolor{black}{can be} improved by the self-consistent-harmonic approximation (SCHA), which is able to correctly foresee the existence of the BKT phase transition. The idea behind the SCHA approximation is to replace the original Hamiltonian with a quadratic one, whose couplings are optimized according to a variational principle. 

In the SR case the natural variational parameter is 
\textcolor{black}{an} effective coupling $\tilde{J}$, which replaces the bare one $J$ in the effective theory. As shown in Refs.\,\cite{POKROVSKY1973,Pires1997}, in this case $\tilde{J}$ jumps discontinuously to zero as the temperature increases, implying that the exponent of the correlations becomes infinite in the high-temperature region. 
This path has been pursued for the case of a power-law interaction as well: in this case both the coupling $J$ and the exponent of the interaction $\sigma$ are replaced by variational parameters $\tilde{J}$ and $s$. The results are unambiguous as long as $\sigma>2$, predicting a phenomenology analogous to the SR case\,\cite{Giachetti2021EPL}. Our aim is to generalize such an analysis 
in order to correctly \textcolor{black}{deal with} the $\sigma \in (0,2]$ regime.

To achieve this scope, we replace the cosine in the original Hamiltonian \eqref{model} with a quadratic term
\begin{equation} \label{ansatz}
\beta H_0 =  \frac{1}{4} \sum_{\mathbf{i},\mathbf{j}} \tilde{J}(\mathbf{r}) ( \theta_{\mathbf{i}} - \theta_{\mathbf{j}})^2,
\end{equation}
$\tilde{J}(r)$ being a completely arbitrary function of $\mathbf{r} = \mathbf{i} - \mathbf{j}$, to be determined \textcolor{black}{from free energy minimization} in a self-consistent way. 

The quadratic Hamiltonian $H_0$ induces the Boltzmann measure
\begin{equation}
\me{\cdot}_0 = \frac{1}{Z_0} \int \prod_{\mathbf{j}} d \theta_{\mathbf{j}} \ e^{- \beta H_0},
\end{equation}
where 
\begin{equation}
Z_0 = e^{- \beta F_0} = \int \prod_{\mathbf{j}} d \theta_{\mathbf{j}} \ e^{- \beta H_0}
\end{equation}
is the partition function of the model. The variational principle establishes that our best ansatz minimizes the variational free energy 
\begin{equation} \label{mathcalF}
\mathcal{F} = \beta F_0 + \beta \me{H}_0 - \beta \me{H_0}_0.
\end{equation}
On the other hand, from the equipartition theorem it follows that $\me{H_0}_0 = \frac{N}{2 \beta}$ is independent on the choice of $\tilde{J}(\mathbf{r})$, so that we can safely ignore it. Since $H_0$ has the same quadratic structure of the low-temperature Hamiltonian, 
we can diagonalize $H_0$ as well by means of the Fourier transform
\begin{equation} \label{diagH0}
\beta H_0 = \frac{1}{2} \sum_{\mathbf{q} \in IBZ} \tilde{K}(\mathbf{q}) |\theta_{\mathbf{q}}|^2,
\end{equation}
where $\tilde{K}(\mathbf{q})$ is given by Eq.\,\eqref{Kq} with $J(r)$ replaced by $\tilde{J}(r)$. As shown in 
Appendix \ref{AppB}, 
the variational free energy takes the form
\begin{equation} \label{Fvar}
\mathcal{F} = \frac{1}{2} \sum_{\mathbf{q} \in IBZ} \ln \tilde{K}(\mathbf{q}) - \frac{N}{2} \sum_{\mathbf{r}} J(r) e^{- \tilde{G}(\mathbf{r})},
\end{equation}
where, once again, $\tilde{G}(\mathbf{r})$ is given by Eq.\,\eqref{Gr} with $J(r)$ replaced by $\tilde{J}(r)$. In order to 
\textcolor{black}{simplify} the notation, in the following we will drop the $\sim$ 
\textcolor{black}{symbol for 
$K(\mathbf{q})$}. 

Since in Eq.\,\eqref{Fvar} $\tilde{J} (\mathbf{r})$ appears only through the $K(\mathbf{q})$, in order to find the minimum is sufficient to derive $\mathcal{F}$ with respect to the latter. By exploiting the fact that
\begin{equation}
\frac{\delta \textcolor{black}{\tilde{G}}(\mathbf{r})}{\delta K(\mathbf{q})}  = - \frac{1}{N} \frac{1 - \cos( \mathbf{q} \cdot \mathbf{r})}{K(\mathbf{q})^2}, 
\end{equation}
we find 
\begin{equation}
\frac{\delta \mathcal{F}}{\delta K(\mathbf{q})}  = \frac{K(\mathbf{q}) - \sum_r J(r) \left( 1 - \cos (\mathbf{q} \cdot \mathbf{r}) \right). e^{- G(\mathbf{r})}}{2 K(\mathbf{q})^2}.
\end{equation}
\textcolor{black}{By using} the definition \eqref{Kq} of $K(q)$ and noticing that the above expression has to be valid for each value of $\mathbf{q} \in IBZ$ \textcolor{black}{(first Brillouin zone)}, we find the condition 
\begin{equation} 
J(r) = \tilde{J}(\mathbf{r}) e^{\textcolor{black}{\tilde{G}}(\mathbf{r})}.
\end{equation}
Since we are interested in the large length-scales regime (i.e in the continuous limit), we can assume $\tilde{J}$ to be a function of $r$ only: in this case, indeed, $\textcolor{black}{\tilde{G}}(\mathbf{r})$ only depends on the modulus $r$ as well, so that the above condition can be written in terms of \textcolor{black}{single} variable functions 
\begin{equation} \label{Min}
J(r) = \tilde{J}(r) e^{\textcolor{black}{\tilde{G}}(r)}.
\end{equation}
Let us notice how, in this limit, one has to redefine $\mathbf{r} \rightarrow a \mathbf{r}$, $\mathbf{q} \rightarrow a^{-1} \mathbf{q} $, $J, \tilde{J} \rightarrow a^{-2} J,  a^{-2} \tilde{J}$. 

Let us now discuss the possible solutions of Eq.\,\eqref{Min}. 
\textcolor{black}{The possible 
asymptotic behaviors of $K(q)$ and $\tilde{G}(q)$ in terms of $\sigma$ are examined in Appendix \ref{AppA}. We find:} 
we find that:
\begin{itemize}
    \item If $\tilde{J}(r)$ decays at infinity faster than $r^{-4}$ \textcolor{black}{(e.g. an exponential or a fast-decaying power law)}, then we have that $\textcolor{black}{\tilde{G}}(r) \sim \ln(r)$ as $r \rightarrow \infty$. Since in this case $e^{\textcolor{black}{\tilde{G}}(r)}$ is a power law, it follows from Eq.\,\eqref{Min} that $\tilde{J}(r)$ must behave asymptotically as a power-law as well. 
    We can then assume \textcolor{black}{that for large $r$} 
    \begin{equation}
    \tilde{J}(r) \sim \tilde{J} r^{-2-s}    
    \end{equation}
    \textcolor{black}{for some $s>2$ to be determined,}  so that 
    \begin{equation} \label{primo}
        \textcolor{black}{\tilde{G}} (r) \sim \eta(\tilde{J}) \ln \frac{r}{a} + A \tilde{J}^{-1},
    \end{equation}
    where $\eta(\tilde{J}) = p \tilde{J}^{-1}$ and $A$, $p$ are non-universal constants. Finally, from Eq.\,\eqref{Min}, we have the conditions
    \begin{equation}
        \sigma = s - \eta(\tilde{J}); \hspace{0.5cm} J = \tilde{J} e^{A/\tilde{J}}. 
    \end{equation}
    In this case the correlation functions decay as $e^{- \textcolor{black}{\tilde{G}} (r)} \sim r^{-\eta(\tilde{J})}$. Then, as long as $\tilde{J}$ is 
    \textcolor{black}{non-vanishing}, we find a quasi-long-range order, characteristic of the BKT phenomenology. 
    \item If the variational coupling behaves as $\tilde{J}(r) \sim \tilde{J} r^{-2 - s}$ with $s \in (0,2)$, then $G(r) = A \tilde{J}^{-1} + O(r^{s-2})$. In this case we then have: 
    \begin{equation}
        \sigma = s; \hspace{0.5cm} J = \tilde{J} e^{A/\tilde{J}}, 
        \label{26}
    \end{equation}
    leading to correlations behaving as $e^{- \textcolor{black}{\tilde{G}} (r)} \sim e^{-A\tilde{J}^{-1}}$. Then as long as $\tilde{J}$ is \textcolor{black}{non-vanishing}, we find a finite magnetization $\sim e^{-A\tilde{J}^{-1}/2}$. 
\end{itemize}
In both cases, the equation for $\tilde{J}$ has the same form 
of the \textcolor{black}{nearest-neighbour} case. 
By introducing $\tilde{J} = A x$, $J = A y$ it can be written as: 
\begin{equation}
    y = x e^{1/x}.
\end{equation}
The minimum of the r.h.s. \textcolor{black}{term} is in correspondence of $x=1$, $y=e$ so that \textcolor{black}{we have two solutions for $J > J_c \equiv e A$, and only the trivial solution for $J < J_c$, signaling a jump in $\tilde{J}$ from $\tilde{J}_c = A$ to zero. However, only the larger of the two solutions present for $J > J_c$ is physically acceptable. Indeed it is the only one to have the \textcolor{black}{correct} asymptotic behavior $\tilde{J}(r) \sim J(r)$ in the large $J$ regime, in which our SCHA becomes the low-temperature approximation studied in Sec. III.} 
The meaning of this low-temperature, finite $\tilde{J}$ phase, however, depends on the whether $s<2$ (ordered phase) or $s>2$ (quasi-long-range-ordered BKT phase). Let us notice that, in the latter case, we have that $\eta = p \tilde{J}^{-1}$ of Eq.\,\eqref{primo} cannot be larger than a given value $\eta_{c} = p \tilde{J}_c^{-1}$. As a consequence we have that:
\begin{itemize}
    \item For $\sigma > 2$, the only possibility is that \textcolor{black}{$s = \sigma + \eta_{SR} > 2$}. We are then in the first case, so that we get the usual BKT phenomenology. This 
    \textcolor{black}{in agreement with the findings of}
    Ref.\,\cite{Giachetti2021EPL}. 
    \item For $\sigma < 2 - \eta_c$, the only possibility is that \textcolor{black}{$s = \sigma$} with $s < 2$. We are then in the second case, i.e. we find a finite magnetization for low temperature.
    \item For $2- \eta_c < \sigma < 2$ both solutions are actually viable so that it is unclear whether the system is in the ordered or in the quasi-long-range ordered phase. To establish which solution should we choose we should compute $\mathcal{F}$ on both solutions. 
    \textcolor{black}{The dependence of $\mathcal{F}$ on the non-universal details of the model, and in particular on the short-range behavior of $\tilde{J}(r)$. hinders the 
   possibility to reach a definite conclusion for this regime within the SCHA. 
   }
\end{itemize}

Although non-conclusive, our self-consistent analysis accounts for the possibility of the existence of an intermediate region, in which the ordered phase or the BKT behavior can prevail, depending on the temperature. Keeping in mind the results of the Berezinskii approximation, it appears sensible to think that the magnetized phase will prevail at lower temperatures, while the quasi-long-range ordered phase will, possibly, correspond to an intermediate range of temperature. Let us notice, however, that the predictions of our analysis are non-universal and, thus, their quantitative outcome depends on the peculiar model under study. Moreover, the first-order phase transition foreseen for $\sigma < 2 - \eta_c$ is 
\textcolor{black}{could} be an artifact of the approximation, since the critical behavior of the model is known to be captured by the mean-field approximation for $\sigma <1$, \textcolor{black}{which foresees a second-order phase transition, see e.g. Ref.\,\cite{defenu2017criticality}}. 

\section{Field-theoretical approach}
\noindent
In order to go beyond the limits of the SCHA, and capture the universal quantities we are interested in, we 
\textcolor{black}{resort} to a field-theoretical approach, introducing a continuous action, which encodes the same physics of our Hamiltonian \eqref{model}. 
First, we write the coupling as $J (r) =
J_{S} (r) + J_{LR} r^{- (2+ \sigma)}$, $J_S$ being a SR term which accounts for the small-distances behavior. This decomposition allows us to refine our low-temperature approximation. This, in fact, is fully justified for the SR coupling, even at intermediate temperatures, since it couples \textcolor{black}{neighboring} sites. At the same time, it can become too crude for the long-range part of the Hamiltonian. Since this couples far-away pairs, there can be smooth configurations for which the phase difference $\theta_{\mathbf{i}}- \theta_{\mathbf{j}}$ is not necessarily small, and these configurations may give a significant contribution to the Hamiltonian in an intermediate range of temperatures. However, even for the SR term, the approximation \emph{a là} Berezinskii is unable to capture the presence of vortices.

\textcolor{black}{To further proceed,} we expand the cosine in the SR part
\begin{equation}
    1 - \cos(\theta_{\mathbf{i}}-\theta_{\mathbf{j}}) \approx \frac{1}{2} \left(\theta_{\mathbf{i}}-\theta_{\mathbf{j}}\right)^2 \approx \frac{1}{2} |\nabla \theta|^2,
\end{equation}
where the last substitution is justified by the fact that in the SR part of the Hamiltonian only \textcolor{black}{neighbouring} lattice sites are important. We then find
\begin{equation} \label{action}
  S[\theta] =  \frac{J}{2} \int d^{2}x |\nabla \theta|^2 + S_{\mathrm LR} [\theta],
\end{equation}
where we introduced the LR perturbation
\begin{equation} \label{rawSLR}
S_{\mathrm LR} = J_{\rm LR} \int d^2 x \int_{r>a} \frac{d^2 r}{r^{2+ \sigma}} [1 - \cos \left(\theta (\mathbf{x}) - \theta (\mathbf{x} + \mathbf{r} ) \right)].
\end{equation}
In turn, this term can be rewritten in terms of the fractional Laplacian, whose definition is given, along with the details of the calculation, in Appendix \ref{AppC}. We obtain
\begin{equation} \label{fractionalS}
S_{\mathrm LR} = \frac{g}{2} \int d^2 x \, e^{-i \theta} \, \nabla^{\sigma} e^{i \theta},
\end{equation}
with $g = J_{\rm LR}/\gamma_{2, \sigma}$ and $\gamma_{2,\sigma} = 2^{\sigma} \Gamma(\scriptstyle \frac{1 + \sigma}{2} \displaystyle) \pi^{-1} |\Gamma(\scriptstyle -\frac{\sigma}{2} \displaystyle)|^{-1}$. This term is intrinsically interacting. Moreover it is invariant under global translations $\theta(\mathbf{x}) \rightarrow \theta(\mathbf{x}) + \alpha$, and it correctly \textcolor{black}{catches} the fact that the field $\theta$ is periodic. However, because of the kinetic term in $S[\theta]$, the whole action still does not correctly describes topological phenomena. 

Let us remind briefly of the $g=0$ case. Here one can introduce vortices 
in Eq.\,\eqref{action} i.e.\, 
\textcolor{black}{evaluating} the energy cost of a 
\textcolor{black}{vortex} configuration, and the core energy cost of a single vortex $\varepsilon_c$, which can be absorbed into the definition of the vortex fugacity $y = \exp(- \varepsilon_c)$. Since topological and non-topological configurations \textcolor{black}{decouple} in the quadratic part of the action, this leads to the Coulomb gas picture and to the well-known Kosterlitz-Thouless RG equations\,\cite{KT}. In turn, this implies that all the fixed points for $y=0$ and $J > \frac{2}{\pi}$, which correspond to a Gaussian massless action, are stable. There, the low-temperature approximation becomes correct (with the original $J(T)$ replaced by the one \textcolor{black}{corresponding} to the fixed point), so that the power-law scaling observed for the two-point functions is recovered. For $J < \frac{2}{\pi}$, however, vortices become relevant and the theory flows towards the disordered regime. 
\begin{figure}
    \centering
    \includegraphics[scale=0.44]{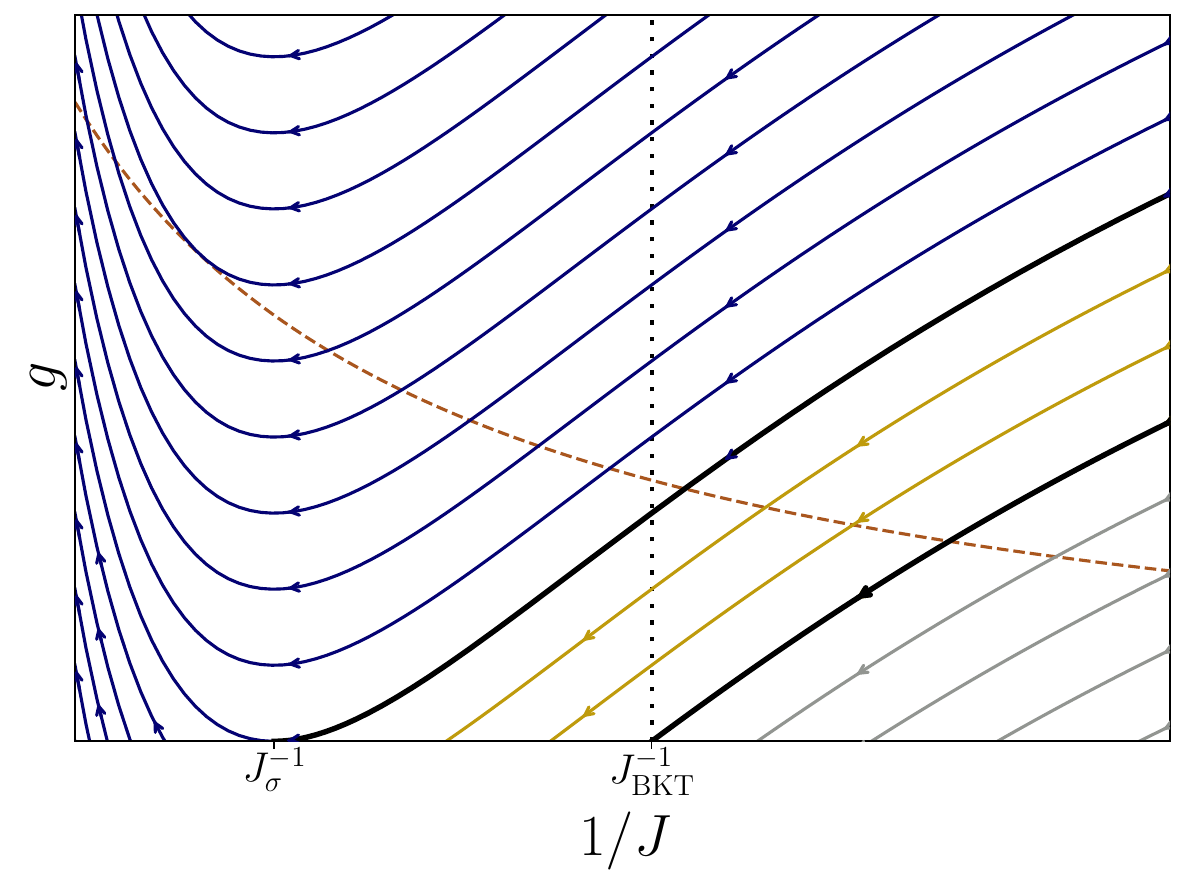}
    \caption{Behavior of the RG flow for $y=0$ in the $\frac{7}{4} < \sigma < 2$ regime. 
    The dotted black line separates the region in which vortices are relevant (
    \textcolor{black}{right}, $\dot{y}_{\ell}>0$), or not (
    \textcolor{black}{left}, $\dot{y}_{\ell}<0$). We find three regions, corresponding to the three phases: for high temperatures (grey), the vortices push the system towards the disordered phase, for intermediate temperatures (gold) the system flows towards a SR $g=0$ fixed point, while for low-temperatures (blue) the LR perturbation becomes relevant, \textcolor{black}{pushing} the system towards an ordered region. In bold black \textcolor{black}{we plot} the separatrices between the different phases.}
    \label{Fig1}
\end{figure}

In order to study the transition between the SR and the LR regime, we can consider the effect of a small $g$ deformation of the Gaussian fixed points. In this case, we expect to be able to parametrize our theory in terms of three parameters: $J$, $g$ and $y$. In Appendix \ref{AppD} we derive the RG flow perturbatively in $y$, $g$. However, we provide here an argument to understand what 
\textcolor{black}{one can expect on general grounds}. First, we notice that the fixed points with $J<\frac{2}{\pi}$ are already unstable under topological perturbation\textcolor{black}{s}, so that we will stick to the stable fixed line with $J>\frac{2}{\pi}$. Here, \textcolor{black}{the} LR perturbation can be relevant or not depending on the scaling dimension
$\Delta_{g}$ of the coupling $g$, $g_{\ell}\approx\exp(\Delta_{g}\ell)$ for $\ell\gg1$,
where $\ell=\ln(r/a)$ is the RG time. 

Due to the quadratic nature of the measure, \textcolor{black}{one has}
\begin{equation} \label{lr_scaling} 
\me{e^{i \textcolor{black}{\big(}\theta(\mathbf{x}) - \theta(\mathbf{x'}) \textcolor{black}{\big)}}} = e^{- \frac{1}{2} \me{\left( \theta(\mathbf{x}) - \theta(\mathbf{x'}) \right)^2}} = |\mathbf{x} - \mathbf{x'}|^{-\eta_{\rm sr}(J)},  
\end{equation}
where $\eta_{\rm sr}(J) = \frac{1}{2 \pi J}$ corresponds to the exponent of the two-point function for $g=0$ \cite{KT,K,Jose1977}. Then 
\begin{align} \label{ScalingDimension}
\Delta_{g}=2-\sigma-\eta_{\mathrm{sr}}(J).
\end{align}
The LR perturbation 
is relevant only in the regime $\sigma < 2-\eta_{\rm sr}(J)$. This bears some similarly with the SSB case\,\cite{defenu2015fixed}, but with the main difference that the anomalous dimension is temperature dependent. 

For $\sigma>2$\textcolor{black}{,} the LR term is always irrelevant. For $\sigma <2$ Eq.\,\eqref{ScalingDimension} predicts that the LR perturbation is always relevant at low enough temperatures. Since the points which are stable under the proliferation of vortices are those with $J> \frac{2}{\pi}$, we have that $0 < \eta_{\rm sr} < \frac{1}{4}$, as in the usual BKT theory (which indeed corresponds to the $g=0$ theory). As a consequence, for $\frac{7}{4} < \sigma < 2$, we have that it exists an intermediate range of value of $J$ for which the Gaussian theory is stable with respect to both the topological and the LR perturbations, leading to conventional quasi-long-range order in a given temperature window. Instead, for $\sigma < 7/4$ the BKT stable fixed line is completely swallowed by the action of the LR perturbation, so that our perturbative picture breaks down. However, it is sensible to assume that in this regime the system simply undergoes an order-disorder phase transition. 

This picture is essentially in agreement with the results of the SCHA. However, our results no longer depends on the regularization procedure or the exact form of $J_{\rm S} (r)$ and are 
\textcolor{black}{genuinely} universal. Our observation can be 
\textcolor{black}{strengthened} by the renormalization equations (see Appendix \ref{AppD} for details) which, at the leading order in $y$ and $g$ are given by
\begin{equation} \label{RG LR}
\begin{split}
\frac{dy_{\ell}}{d\ell} &= (2 - \pi J) y, \\
\frac{d g_{\ell}}{d\ell} &=\Big( 2 - \sigma - \frac{1}{2 \pi J_{\ell}} \Big) g_{\ell},  \\
\frac{dJ_{\ell}}{d\ell} &= c_{\sigma} \eta_{\rm sr}(J_{\ell}) g_{\ell}, 
\end{split}
\end{equation}
with $c_{\sigma} = \frac{\pi}{2} a^{2-\sigma} \int^{\infty}_1 du \ u^{1 - \sigma} \mathcal{J}_{0}
(2 \pi u)$ and $\mathcal{J}_{0} (x)$ is the zeroth-order Bessel function of the first kind. 
\textcolor{black}{Our} argument correctly predicts the first two equations, but not the third one, i.e. the renormalization of  the spin-waves stiffness. In particular, for $7/4 < \sigma < 2$\textcolor{black}{,} we find a line of stable quadratic fixed points for $g=y=0$ and 
\textcolor{black}{$J_{\rm BKT} \equiv \frac{2}{\pi} < J < J_{\sigma} \equiv \frac{1}{2 \pi (2 -\sigma)}$}, as expected. 
\textcolor{black}{The} behavior of $y_{\ell}$, at this order, is completely determined by $J$ and, as long as $J > J_{\rm BKT}$, $y_{\ell} \rightarrow 0$.

Then, for $7/4 < \sigma <2$ we can characterize the transition between the ordered phase and the quasi-long-range ordered one by looking at the $y=0$ plane. As long as 
\textcolor{black}{$g$} is small, we can explicitly identify the form of the flow trajectories of Eqs.\,\eqref{RG LR}: 
\begin{equation} \label{flux}
g_{\ell} (J) =  \frac{\pi (2 - \sigma)}{ c_{\sigma}} \left[ \left( J_{\ell} - J_{\sigma} \right)^2 + k \right].
\end{equation}
If $k<0$ the trajectory arrives at the fixed point line $g=0$ for some $J < J_{\sigma}$, while for $k>0$ the trajectory starts from the $g=0$ line for $J > J_{\sigma}$ and it goes to infinity, signaling the existence of a new, low-temperature, phase. The separatrix is given by the semi-parabola with $k=0$, $J \leq J_{\sigma}$. The graphical depiction of the RG flow of Eqs.\,\eqref{RG LR}, in the $y=0$ plane, 
\textcolor{black}{is} shown in Fig.\,\ref{Fig1}. 

Being Eqs.\,\eqref{RG LR} a perturbative result we derived for small $g_{\ell}$ and $y_{\ell}$, its use in the low temperature region ($T<T_c$) is not justified, as $g_{\ell}$ grows indefinitely. 
\textcolor{black}{H}owever, for $T \rightarrow T_c^{-}$, the \textcolor{black}{amount of} time spent by \textcolor{black}{the} flow in the region close \textcolor{black}{to} $J=J_{\sigma}$, $g=0$ becomes larger and larger, so that the scaling behavior of $g_{\ell}$ with $T$ in this regime can be reliably obtained from Eqs.\,\eqref{RG LR}. In particular, for fixed $\ell$ such that $J_{\ell} > J_{\sigma}$ we have 
\begin{equation} \label{gscaling}
    g_{\ell} \sim e^{-B(T-T_c)^{-1/2}}
\end{equation}
where $B$ is a non universal constant (see Appendix 
\ref{AppE} for details).

Although the infrared regime is beyond \textcolor{black}{the} reach of our perturbative analysis, it is possible to guess the corresponding form taken by the action on physical grounds. Indeed, the coupling $J$ diverges there, suppressing spatial fluctuations of the phase $\theta(\mathbf{x})$, as confirmed by the rigorous results of Ref.\,\cite{kunz1976first}, which predict a finite magnetization for low enough temperatures. \textcolor{black}{This} suggests that, in the infrared region corresponding to the low-temperature phase, we are allowed to Taylor expand the exponential in Eq.\,\eqref{action}, so that the action becomes
\begin{equation} \label{Gaussian}
S_{\rm LR} = - \frac{g}{2} \int d^2 \mathbf{x} \ \theta \nabla^{\sigma} \theta,
\end{equation}
where we absorbed in $g$ some immaterial constants. A further evidence in favour of this action comes from the fact that, in the limit $J \rightarrow \infty$, we have $\Delta_g = 2 - \sigma - \eta_{sr} (J) \rightarrow 2 - \sigma$, which is exactly the scaling dimension of $g$ in Eq.\,\eqref{Gaussian}. This also suggests that in this regime topological excitations are suppressed. 

Physically speaking, this is due to the fact that, as $J \rightarrow \infty$, the energy cost of highly non-local excitations like the topological one becomes higher and higher, and the presence of relevant LR perturbation further contributes to this.   
\begin{figure}
    \centering
    \includegraphics[width=\linewidth]{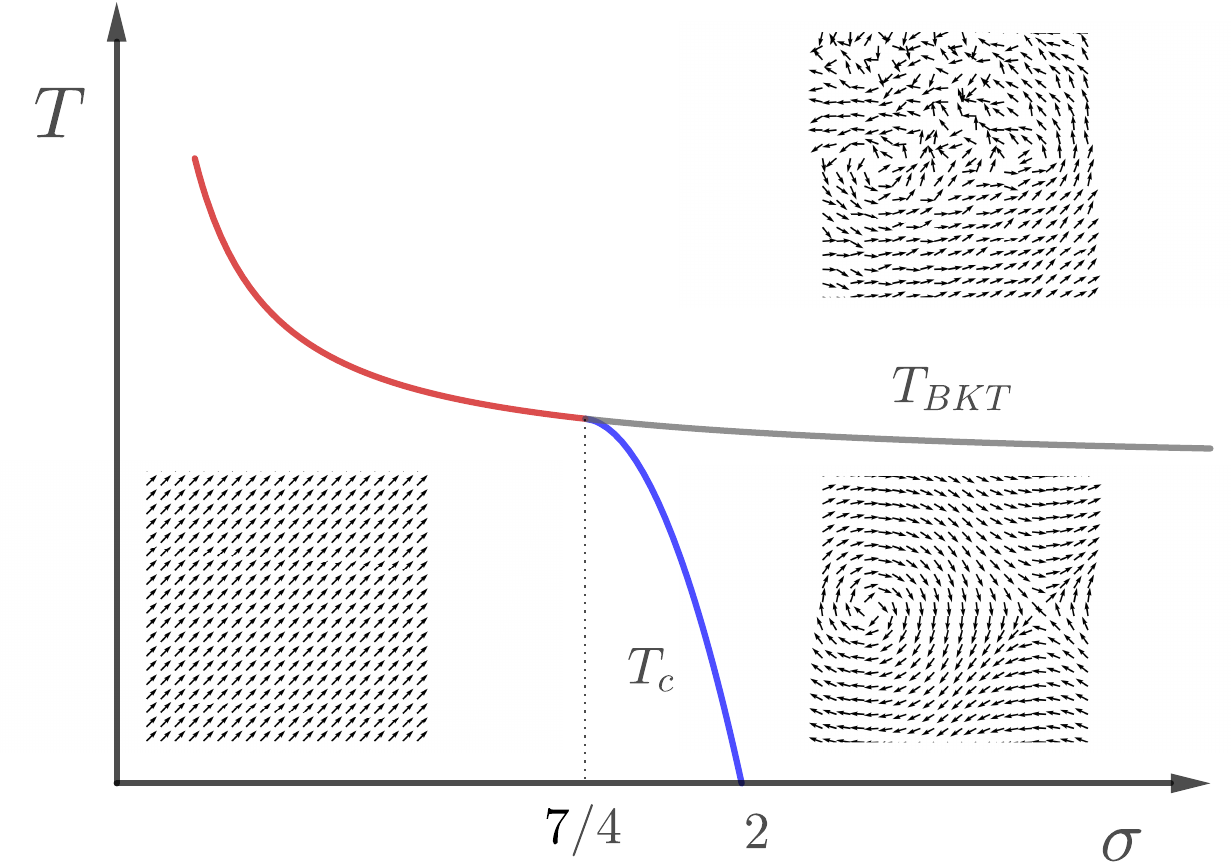}
    \caption{Qualitative phase diagram for the long-range $XY$ model in $d=2$, in which the three phases of the model (ordered, quasi-long-range-ordered, disordered) are shown. For $\sigma>2$ the system undergoes the usual $BKT$ transition (gray line). The value $\sigma^{*}$ at which the LR term becomes relevant is here a function of the temperature and varies from $\sigma = 2$ (at $T=0$) to $7/4$ (when it met the BKT transition temperature). In this range of $\sigma$ then, as the temperature varies, beyond the $BKT$ transition, we find \textcolor{black}{an} infinite order symmetry-breaking transition (blue line). For $\sigma < 7/4$ the quasi-long-range ordered phase disappears so that we only have an order-disorder transition.}
    \label{Fig2}
\end{figure}
Let us notice that the action of Eq.\,\eqref{Gaussian} is nothing but the continuous version of the approximation \emph{a lá} Berezinskii. This tells us that, for small enough temperature, the approximation is indeed reliable. In particular, we can use Eq.\,\eqref{mBer} to derive the magnetization
\begin{equation}
    \ln m \sim g^{-1} \int q^{-\sigma} d^2 \mathbf{q}
\end{equation}
Finally, from Eq.\,\eqref{gscaling} we find the scaling of the magnetization 
\begin{equation} \label{scaling}
\ln m \sim -e^{B(T_c - T)^{-1/2}}
\end{equation}
Here, the phase transition is of infinite order as all the derivatives of the order parameter with respect to the temperature vanish at $T=T_c$. \textcolor{black}{The same can be said for the free energy which would as well exhibit an essential singularity. A similar behavior in $T$ is found approaching $T_{BKT}$ from above, so that this seems to be a general property of BKT phase. The presence of a SSB phase is, however, proper of the LR regime. A similar, albeit not identical, coexistence of finite order parameter and infinite order scaling signatures has been also observed in other long-range statistical mechanics model\textcolor{black}{s}\,\cite{anderson1969exact,thouless1969long, dyson1971ising, cardy1981one, aizeman1988discontinuity}, see also Ref.\cite{barma2019fluctuation} for more examples in this direction. } 

As $\sigma \rightarrow 7/4^{+}$, $T_c$ reaches $T_{\rm BKT}$ from below, leaving only a SSB phase transition. Unfortunately however, our set of equations\,\eqref{RG LR} \textcolor{black}{is} not reliable in this regime: the RG flow \textcolor{black}{spends a considerable} close to the $g=0$, $J = J_{\sigma}$ fixed point, which, in this regime, corresponds to $\dot{y}_{\ell} >0$. As a consequence, $y_{\ell}$ will not remain small. 

Summarizing, when $7/4<\sigma<2$, we find \textit{i)} for $T<T_c$, finite magnetization (ordered phase); \textit{ii)} for $T_c<T< T_{\rm BKT}$ quasi-long-range-order with zero magnetization and temperature-dependent power-law decay in the two-point correlation function (BKT phase); \textit{iii)} for $T>T_{\rm BKT}$ zero magnetization (disordered phase). The system, due to the power-law character of the interactions, exhibit\textcolor{black}{s} power-law decaying two-point functions, also in the high-temperature phase, although with the same exponent of the coupling $\me{\mathbf{S}(\mathbf{r}) \cdot \mathbf{S} (0) } \sim r^{-2-\sigma}$\,  \cite{spohn1999decay,kargol2005decay,kargol2014decay}. The qualitative form of the phase diagram of the model is shown in Fig.\,\ref{Fig2}. 

\section{Villain approximation}
\noindent
The study of the SR XY model in $d=2$ is greatly simplified by the possibility of performing the so-called Villain approximation, in which we replace in the Hamiltonian in Eq.\,\eqref{model_xy}
\begin{equation}
    1 - \cos(\theta_{\mathbf{i}}-\theta_{\mathbf{j}}) \rightarrow \frac{1}{2} \left(\theta_{\mathbf{i}}-\theta_{\mathbf{j}} - 2 \pi n_{\mathbf{i},\mathbf{j}} \right)^2
\end{equation}
where $n_{\mathbf{i},\mathbf{j}}$ is an auxiliary link variable which assumes integer values and which has to be traced out. Although the variable appears only quadratically, the model exhibits a BKT phenomenology, since it correctly reproduces the $\theta_{\mathbf{i}} \rightarrow \theta_{\mathbf{i}} + 2 \pi m$ symmetry with $m \in \mathbb{Z}$. In particular, by integrating out the $\theta_{\mathbf{i}}$, the Villain Hamiltonian can be exactly mapped into the Coulomb gas Hamiltonian (in any dimension).

While the same approximation could be, in principle, performed even for the case of long-range-decaying couplings, it is not guaranteed for the resulting phenomenology to be in the same universality class of the LR $XY$ model. There are, in fact, reasons to believe this is not true, as we are going to argue. First, let us notice that, if on the one hand \textcolor{black}{the} Villain approximation improves \textcolor{black}{Berezinskii's} result by taking into account the vortices, on the other hand, it can only account for interaction between topological excitations which is quadratic in their charges. As already seen in the context of the field theory, however, such a quadratic approximation is not always justified. 

To be more concrete, let us express the action $S[\theta]$ of the $XY$ model in terms of the topological charges. We start by splitting the field in a topological and a non-topological spin-waves part:
\begin{equation} \label{decompos}
 \theta(\mathbf{x}) = \theta_{\rm 0} (\mathbf{x}) + \theta_{\rm top} (\mathbf{x})  
\end{equation}
where, for every closed circuit,
\begin{equation}
    \oint \nabla \theta_0 \cdot d \mathbf{x} = 0 \hspace{0.5cm} \oint \nabla \theta_{\rm top} \cdot d \mathbf{x} = 2 \pi m_{\rm enc}
\end{equation}
$m_{\rm enc}$ being the total topological charge enclosed in the circuit. The ambiguity in the decomposition can be lifted if we impose the further constraint $\nabla \cdot \nabla \theta_{\rm top} = 0$. This, in turn, allows us to write $\theta_{\rm top}$ in terms of the vortex configuration so that we can finally write Eq.\,\eqref{action} as
\begin{equation}\label{top+ntop}
\begin{split}
    S[\theta]  &= \frac{J}{2} \int d^2 \mathbf{x} \ |\nabla \theta_0|^2 - \pi J \sum_{j \neq k} m_j m_k \ln |\mathbf{x}_k - \mathbf{x}_j| \\ &+ \varepsilon_c \sum_k m_k^2 + \frac{g}{2} \int d^2 x \, e^{-i (\theta_0 + \theta_{\rm top}) } \, \nabla^{\sigma} e^{i (\theta_0 + \theta_{\rm top})}
    \end{split}
\end{equation}
where $m_k$ and $\mathbf{x}_k$ are respectively the charges and the positions of each vortex (see Appendix 
\ref{AppF} for details). For small values of $g$ we can integrate out the non-topological component of the field, $\theta_0$, finding
\begin{equation} \label{pert}
\begin{split}
    S_{\rm eff} \sim &- \pi J \sum_{j \neq k} m_j m_k \ln |\mathbf{x}_k - \mathbf{x}_j| + \varepsilon_c \sum_k m_k^2 \\ &+ \frac{g}{2} \int d^2 \mathbf{x} \, e^{-i \theta_{\rm top}}  \nabla^{\sigma + \eta_{\rm sr} (J)} e^{i \theta_{\rm top}} 
\end{split}
\end{equation}
where it is implied that, if $\sigma + \eta_{\rm sr} (J) >2$, the above operator has to be interpreted as the usual Laplacian (see Appendix 
\ref{AppF}). In the latter case, in particular, 
\begin{equation}
    -e^{-i \theta_{\rm top}}  \nabla^{2} e^{i \theta_{\rm top}} = |\nabla e^{i \theta_{\rm top}}| = |\nabla \theta_{\rm \textcolor{black}{top}}|^2 
\end{equation}
so that, by replacing $\theta_{\rm top}$ with its expression in terms of the $m_i$, we recover the usual Coulomb gas interaction 
\begin{equation}
\begin{split}
    S_{\rm eff} &\sim - \pi (J+g) \sum_{j \neq k} m_j m_k \ln |\mathbf{x}_k - \mathbf{x}_j| \\
    &+ \varepsilon_c \sum_k m_k^2
\end{split}
\end{equation}
We thus recovered the main result of our analysis, namely, the fact that the system behaves as its SR counterpart even for $\sigma<2$, provided that $\sigma + \eta_{\rm sr} (J) >2$. 

If $\sigma + \eta_{\rm sr} (J) <2$, it is not possible to derive a simple charge-charge interaction, unless we expand the exponential. This expansion is actually not easily justified, since the topological configurations $\theta_{\rm top}$ are spatially extended. This suggests that the higher-order terms in $\theta_{\rm top}$ in the expansion are relevant, so that we have to keep in the Hamiltonian higher-order interaction terms in the $m_k$ (e.g. proportional to $m_i m_j m_k m_p$), which cannot be obtained within the Villain approximation.

Moreover, let us notice that $\sigma + \eta_{\rm sr} (J) <2$ is precisely the condition under which the LR term becomes relevant. In this regime, our field-theoretical analysis foresees a boundless growth for the coupling $g$, so that the integration on $\theta_0$ cannot be performed perturbatively as we did in Eq.\,\eqref{pert}. The fact that spin-waves and topological contributions \textcolor{black}{do} not decouple is \textcolor{black}{a} further proof that the long-range phenomenology cannot be  captured by a quadratic model in $\theta$, as the Villain model. 

\begin{figure}
    \centering
    \includegraphics[width=\linewidth]{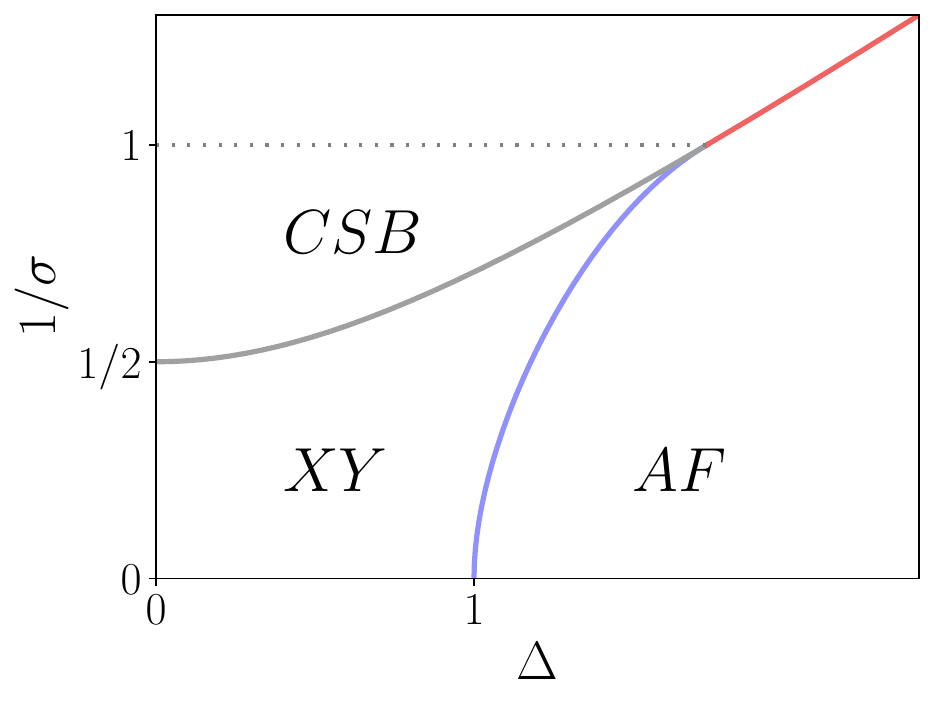}
    \caption{Qualitative phase diagram of the LR $XXZ$ quantum chain studied in Ref.\,\cite{maghrebi2017continuous}. \textcolor{black}{The CSB (continuous simmetry breaking) phase corresponds to the SSB phase of the LR $XY$ model, the AF (anti-ferromagnetic) phase to the disordered one, the gapless $XY$ phase to the BKT phase.} The topology of the phase diagram is the same of the LR $XY$ case (Fig.\,\ref{Fig2}), but here the border between the new broken phase and the \textcolor{black}{gapless} $XY$ phase of the chain goes from $\sigma =2$ to $\sigma=1$, as discussed in the text. In the region $\sigma <1$ the intermediate phase disappears.}
    \label{Fig3}
\end{figure}

\section{Quantum long-range \texorpdfstring{$XXZ$}{XXZ} chain}
\noindent
The mapping between the nearest-neighbors $2d$ classical $XY$ model and the nearest-neighbors, spin-$1/2$, $1d$ quantum $XXZ$ model relies on the local nature of the couplings \cite{Mattis1984}. However, the extension of the equivalence in presence of LR couplings is not obvious. Indeed, the original derivation 
in Ref.\,\cite{Mattis1984},  neglects the 
\textcolor{black}{$z$-$z$}
interaction terms with a range $\ge 4$ lattice sites in the resulting Hamiltonian and introduces a suitable, averaged\textcolor{black}{,}
interaction term for distances lower than three lattice sites. This allows \textcolor{black}{introducing} a bosonic hard-core
condition and the mapping onto the $1d$ quantum $XXZ$ Hamiltonian. This approach 
cannot be straightforwardly applied to the case of the $2d$ $XY$ model with LR interactions, as
one should show the RG irrelevance of terms violating the hard-core condition. 

Therefore, an interesting question is to ascertain 
whether and how the $T=0$ phase diagram of the $1d$ $XXZ$ Hamiltonian with LR interactions
\begin{equation} \label{HamXXZ}
    H = - \sum_{i,r} \frac{1}{r^{1+\sigma}} \left( S^{x}_{i} S^{x}_{i+r} + S^{y}_{i} S^{y}_{i+r} - \Delta S^{z}_{i} S^{z}_{i+r}  \right)
\end{equation}
(where $\sigma>0$ and $S^{x,y,z}$ are the components of the spin $1/2$ operators) is related to that of the $2d$ classical $XY$ model. It is worth noting that Hamiltonian \eqref{HamXXZ} has been studied in Ref.\,\cite{maghrebi2017continuous} through the bosonization technique and numerical simulations. The effective action derived there describing the $XXZ$ model bears a strong resemblance to the one of Eq.\,\eqref{action}. The resulting RG flow is therefore similar to the one of Eq.\,\eqref{RG LR} indicating that the LR-SR crossover of the two model\textcolor{black}{s} is somehow analogous. In particular, identifying in Eqs. (4) and (8) of Ref.\,\cite{maghrebi2017continuous} $K,g_{\rm LR},g$ with $\pi J/4, g, y$ respectively, these flow equations becomes
\begin{equation} \label{RG XXZ}
\begin{split}
\frac{dy_{\ell}}{d\ell} &= (2 - \pi J) y \\
\frac{d g_{\ell}}{d\ell} &=\Big( 2 - \sigma - \frac{2}{\pi J_{\ell}} \Big) g_{\ell}  \\
\frac{dJ_{\ell}}{d\ell} &= A(J) g_{\ell} 
\end{split}
\end{equation}
$A(J)$ being a function of $J$, depending on the RG-scheme. The similarity between Eq.\,\eqref{RG LR} and Eq.\,\eqref{RG XXZ} can be traced back both to the fact that the 
\textcolor{black}{$z$-$z$} LR interaction is actually irrelevant (for $\sigma>0$) and to the fact that the coefficient of the time derivative in the bosonized action does not renormalize (see the corresponding discussion for $\sigma<0$ in Ref.\,\cite{Botzung2021}). This also explains why both phase diagrams feature three phases: the SSB, the disorder and the BKT one, 
\textcolor{black}{which are called in Ref.\,\cite{maghrebi2017continuous} continuous symmetry breaking (CSB) phase, anti-ferromagnetic (AF) phase and gapless $XY$ phase, respectively.}

\textcolor{black}{Let us notice that in the LR regime, the usual mapping based on the coherent spin states, would result in a highly anisotropic action, due to the fact that one still has SR interactions along the Euclidean time direction \cite{defenu2016anisotropic}. In principle, a similar anisotropy would be present in the bosonized action of Ref.\,\cite{maghrebi2017continuous} due to the presence of the $z$-$z$ LR contribution. As already noticed, however, this term turns out to be irrelevant so that the effective infrared action is actually isotropic.
}

Notice that Eqs.\,\eqref{RG XXZ} predict that the BKT phase extends here up to $\sigma = 1$, rather than $\sigma = 7/4$. This difference boils down to the fact that, in the SR regime, in correspondence to the border between the $XY$ and the anti-ferromagnetic phase ($\Delta = 1$)\textcolor{black}{,} the correlation functions of the $x$,$y$ components decay as $r^{-1}$, rather than $r^{-1/4}$, so that the corresponding anomalous dimension is $\eta_{\rm sr} = 1$ and therefore, using the formalism introduced with \textcolor{black}{Sak's} criterion, the smallest value of $\sigma_*$ is $1$ and $\sigma_*$ ranges between $1$ and $2$. This has to be compared with the range for $\sigma_*$ between $7/4$ and $1$ for the classical LR $XY$ at finite temperature in $d=2$.

Even if the ranges of $\sigma_{*}$ do not coincide, the qualitative form of the phase diagrams of the two models is, therefore, the same, as it is possible to see from Fig.\,\ref{Fig3}. This is similar to the results in Fig.\,1 of Ref.\,\cite{maghrebi2017continuous} \textcolor{black}{(let us notice that $1/\alpha \equiv 1/(1+\sigma)$ on the vertical axis is replaced by $1/\sigma$  in our Fig.\,\ref{Fig3})}. In that figure 
two lines, one separating the BKT and the SSB phases, and the other separating the BKT and the disordered phases, are drawn as extracted from numerical DMRG simulations. It is not clear by the figure whether and where the two lines are going to merge in correspondence of $\sigma = 1$, as foreseen by Eqs.\,\eqref{RG XXZ} and in agreement with the picture presented in this paper. Therefore, more extensive numerical simulations are needed to confirm the predictions from Eqs.\,\eqref{RG XXZ} in the vicinity of $\sigma=1$.

Moreover, the substantial similarity between the field theory descriptions of the two models implies that the 
magnetization 
\textcolor{black}{$m_{\parallel}$} in the 
\textcolor{black}{$x$-$y$} plane of the $XXZ$ chain, scales as 


\textcolor{black}{\begin{equation}
    \ln m_{\parallel} \sim -e^{B(\Delta_c-\Delta)^{-1/2}}
\end{equation}}
i.e. it follows the same scaling of Eq\,\eqref{scaling}. This is, to the best of our knowledge, a new prediction for the $XXZ$ LR model.

The present result fits within the effective-dimensionality picture described in Ref.\,\cite{defenu2017criticality}, for LR quantum systems. In particular, the crossover between various LR regimes of the quantum $O(n)$ model in dimension $d$ can be described by introducing the effective dimension 
\begin{equation} \label{effdim}
    D_{\rm eff} = \frac{2 - \eta_{\rm sr}(D_{\rm eff})}{\sigma} d + 1
\end{equation}
valid for any $\sigma<2-\eta_{\rm sr} (d)$, $\eta_{\rm sr} (D)$ being the anomalous dimension of the $D$-dimensional action which describes the SR version of our model. For the case the LR $d=1$ $XXZ$ phase-diagram, when $D_{\rm eff} =2$, we know that the system undergoes a BKT transition, for which $\eta_{\rm sr}$ is not defined. If, however, we take $\eta_{\rm sr} \in [0,1]$, rather than a single value, we find, from Eq.\,\eqref{effdim} we have
\begin{equation}
    2 > \sigma > 1
\end{equation}
that is, exactly the range of coexistence of the SSB and the BKT phases. 

For $\sigma < 2/3$\textcolor{black}{,} we find $D_{\rm eff} > 4$ (with $\eta_{\rm sr} (D_{\rm eff}) = 0$), so that the order-disorder transition is expected to be captured by the mean-field picture. The same thing it is supposed to happen for the $d=2$ classical $XY$ model in the $\sigma < 1$ region. Therefore one could expect that the critical behavior of the quantum chain in the $2/3 < \sigma < 1$ interval can be related to the one of the classical model for  $1 < \sigma < 7/4$. However, given that Eqs.\,\eqref{RG XXZ} and Eq.\,\eqref{RG LR} do not, respectively, apply in these regimes, further studies would be necessary to clarify this point. A table summarizing the analogous phases of \textcolor{black}{the} quantum $XXZ$ and \textcolor{black}{the} classical $XY$ model is presented, see Tab.\,\ref{tab1}.

\textcolor{black}{The present analysis and, especially, the extension of the SCHA approach to LR couplings, discussed in Sec.\,\ref{sec_scha}, 
could be extended to study generic quantum spin-$S$ systems with LR interactions. According to the Haldane conjecture \cite{haldane1983continuum,haldane1983nonlinear,affleck1986proof} antiferromagnetic half-integer spin chains have a gapless excitation spectrum and they are known to display BKT scaling (e.g., for spin-$1/2$ $XXZ$ model in the SR regime at $\Delta=1$), so that we expect the generalization of our picture to higher, half-integer, values of the spin to be straightforward.}

\textcolor{black}{Regarding the integer-spin case, instead, one can have BKT transitions on the ferromagnetic side, as shown for the spin-$1$ chain in \cite{Sanctuary2003}: therefore, one could also expect a phase diagram like the one in Fig.\,\ref{Fig3}, also for integer spin chains; more investigation is, anyway, needed. The application of our studies to ferromagnetic couplings requires the generalization of the transfer-matrix approach beyond the antiferromagnetic case\,\cite{Mattis1984}. In this perspective, one can also resort to SCHA in order to reduce the quantum system to an effective classical one, which can then be simulated via Monte Carlo\,\cite{cuccoli1995effective,cuccoli2006phase}. }

\setlength{\extrarowheight}{0.15cm}
\begin{table}
\centering
\begin{tabular}{|c||c|} 
    \hline
    \toprule
    {$2d$ $XY$} & {$1d$ $XXZ$}  \\
    \hline
    \midrule
    $\sigma<1$  & $\sigma<2/3$ \\
    $1<\sigma<7/4$  & $2/3<\sigma<1$  \\
    $7/4<\sigma<2$  & $1<\sigma<2$ \\
    $\sigma>2$ & $\sigma>2$  \\\bottomrule
        \hline
\end{tabular}
\caption{Correspondence between the phases of the $2d$ $XY$ LR model at finite temperature (left) and in the $1d$ $XXZ$ LR model at zero temperature (right). For the values of $\sigma$ corresponding to the first line we have a mean-field SSB transition; in correspondence of the second line an interacting SSB transition; in correspondence of the third we have both the order-disorder phase transition and the BKT one; in correspondence of the fourth, finally, only the BKT transition is present.\label{tab1}}
\end{table}

\section{Conclusions}
\noindent
We have shown how the rich and unusual phase diagram of the $2d$ classical $XY$ model in presence of long-range (LR) interaction\textcolor{black}{s} can be obtained both  
\textcolor{black}{using a self-consistent approach} and in a field-theoretical way. We also showed how this phase diagram can be related to the one of the LR $XXZ$ quantum chain.

While the mapping between the two models is well established in the nearest-neighbors case\,\cite{Mattis1984}, the similarity between the phase diagrams of the classical and quantum models in the LR case is remarkable.
We provided 
reasons to believe that our results cannot be obtained via the Coulomb gas mapping or via the Villain approximation, as in the case of SR interactions, so that the latter models appear to be in a different universality \textcolor{black}{class} with respect to the $2d$ XY model when LR interactions are present. Along the same lines, numerical results, obtained in Refs.\,\cite{leuzzi2013,cescatti2019analysis}, seem to indicate that the diluted version of the LR $d=2$ $XY$ model has a different phase diagram as well, although further investigation would be needed in order to confirm this expectation. 

We were also able to use the results obtained for the classical $2d$ system to foresee a new, non-analytic, scaling for the order parameter of the LR $XXZ$ chain, which eludes the current classification\,\cite{raju2019normal}. 
\textcolor{black}{Comments on higher-spin quantum chains with LR couplings were also provided.}


The appearance of exotic scaling phenomena by the addition of complex interaction patterns in two-dimensional systems is not unique. Indeed, several interesting interplays between topological defect unbinding and other critical phenomena have been described, e.g. in the case of coupled $XY$ planes\,\cite{bighin2019berezinskii}, $2d$ systems with anisotropic dipolar interactions\,\cite{Maier2004,Fischer2014},  high-dimensional systems with Lifshitz
criticality\,\cite{jacobs1983self,defenu2020topological}, and anisotropic $3d$ XY model\,\cite{shenoy1995anisotropic}.

Let us notice that we cannot derive any prediction from our analysis of the $2d$ $XY$ model in the region $\sigma < 7/4$ ($\sigma<1$ for the case of the $XXZ$ quantum chain). 
\textcolor{black}{It would be important to extend our analysis to the $\sigma < 7/4$ region, in order to determine whether the transition is a second-order one, as one would expect, and also} whether the quantum to classical mapping we established is still valid.

Finally, it would be interesting to study the LR Villain model, in order to establish the failure of the Villain approximation to reproduce the phase diagram of the $2d$ LR $XY$ universality.


\appendix

\section{$K(q)$ and $G(r)$}
\label{AppA}
\noindent
We 
derive the asymptotic expression for $K(q)$ and $G(r)$, presented in the text. In the following we will consider 
\textcolor{black}{a $J(r)$ with a general $\sigma$}, so that our results can be applied to the self-consistent-harmonic 
\textcolor{black}{calculation} as well.

First, we notice that the continuous definition of $G(r)$ and $K(q)$, given in Eq.\,\eqref{Grc} and Eq.\,\eqref{Kqc} respectively, can be expressed, once we integrate on the solid angle, as
\begin{equation} 
    K(q) = 2 \pi \int_{a}^{\infty} dr \ r J(r)  \left(1- \mathcal{J}_0 (qr) \right)
\end{equation}
and 
\begin{equation} 
    G(r) = a^2 \int_{0}^{\Lambda} \frac{dq}{2 \pi}  \frac{q(1 - \mathcal{J}_0 (qr))}{K(q)},
\end{equation}
where $\mathcal{J}_0 (x)$ is the zeroth-order Bessel function of the first kind. 
\textcolor{black}{As} long as $\tilde{J}(r)$ decays slower than $r^{-4}$, then $K^{\prime \prime} (0)$ is finite, so that $K(q) \sim q^2$, the proportionality constant being non-universal. As a consequence, in this case
\begin{equation}
    G(r) \sim \int_{0}^{\Lambda} dq  \frac{1 - \mathcal{J}_0 (qr)}{q},
\end{equation}
which can be rewritten, through the substitution $x = qr$, as
\begin{equation}
    G(r) \sim \int_{0}^{\Lambda r} dx  \frac{1 - \mathcal{J}_0 (x)}{x}.
\end{equation}
For large $r$, the dominant term in the integral above is $\ln (\Lambda r)$. We then conclude that: 
\begin{equation}
    G(r) \sim \eta \ln \frac{r}{a} + B, 
\end{equation}
$\eta$ and $B$ being two non-universal, cutoff dependent, constants. In particular, if we suppose $J(r)$ behaves at infinity as $J r^{-2-\sigma}$, with $\sigma>2$, we have 
\textcolor{black}{$\eta, B \propto J^{-1}$}, so that we can write
\begin{equation}
    G(r) \sim \eta(J) \ln \frac{r}{a} + A J^{-1}
\end{equation}
with $\eta(J) = p/J$ 
\textcolor{black}{for} some $p$.

Let us now consider instead the case $J(r) \sim \frac{J}{r^{2+\sigma}}$, with $\sigma \in (0,2)$. By operating the substitution $x = qr$ we obtain
\begin{equation} 
\begin{split}
    K(q) &= 2 \pi J q^{\textcolor{black}{\sigma}}  \int_{aq}^{\infty} dx \ x^{-2-\sigma}  \left(1- \mathcal{J}_0 (x) \right) \\
    &= 2 \pi J q^{\textcolor{black}{\sigma}} \int_{0}^{\infty} dx \ x^{-2-\sigma}  \left(1- \mathcal{J}_0 (x) \right) + O(q^2),
\end{split}
\end{equation}
so that we have 
\textcolor{black}{$K(q) = c_{\sigma} J q^{\sigma} + O(q^2)$} with $c_{\sigma} = 2^{-\sigma} \pi |\Gamma(-\sigma/2)|/\Gamma(1 + \sigma/2)$. The expression for $G(r)$ becomes then
\begin{equation} 
    G(r) \sim \frac{1}{J} \int_{0}^{\Lambda} dq \ q^{1-\sigma} \left( 1 - \mathcal{J}_0 (qr) \right)
\end{equation}
which asymptotically goes as 
\begin{equation}
    G(r) = A J^{-1} + O(r^{\textcolor{black}{\sigma}-2}). 
\end{equation}

\section{Variational free energy}
\label{AppB}
\noindent
We 
derive \textcolor{black}{in this Appendix} Eq.\,\eqref{Fvar}, starting from the definition of $\mathcal{F}$ in Eq.\,\eqref{mathcalF}. 
\textcolor{black}{Since} $\me{\cos A}_0 = e^{- \frac{1}{2} \me{A^{\textcolor{black}{2}}}_0}$, as valid for every Gaussian measure, we find
\begin{equation}
\begin{split}
\beta \me{H}_0 &= - \frac{1}{2} \sum_{\mathbf{i},\mathbf{j}} J(r) \me{\cos ( \theta_{\mathbf{i}} - \theta_{\mathbf{j}})}_0 \\ &= - \frac{1}{2} \sum_{\mathbf{i},\mathbf{j}} J(r) e^{- \frac{1}{2}\me{( \theta_{\mathbf{j}} - \theta_{\mathbf{j}})^2}_0} \\ &= - \frac{N}{2} \sum_{\mathbf{r}} J(r) e^{- G(|\mathbf{j}-\mathbf{i}|)} \\ &= - \frac{N}{2} \sum_{\mathbf{r}} J(r) e^{- G(\mathbf{r})}, 
\end{split}
\end{equation}
where we used the definition \eqref{Gr} of $G(\mathbf{r})$. By the 
\textcolor{black}{diagonalized} form \eqref{diagH0} of $H_0$, it follows immediately
\begin{equation}
\beta F_0 = \frac{1}{2} \sum_{\mathbf{q} \in IBZ} \ln K(\mathbf{q}). 
\end{equation}
Plugging this 
\textcolor{black}{result} into the expression for $\mathcal{F}$, we find the desired result. 

\section{Fractional Laplacian and $S_{\rm LR}$}
\label{AppC}
\noindent
We want now to provide the definition of the fractional Laplacian, that we used, and derive the form \eqref{fractionalS} of $S_{\rm LR}$. 

For any $\sigma \in (0,2)$, and a function $f(\mathbf{x}):  \, \mathbb{R}^d \rightarrow \mathbb{R}$ one can define $\nabla^{\sigma} f(\mathbf{x})$  as:
\begin{equation} \label{frac}
\nabla^{\sigma} f(\mathbf{x}) \equiv \gamma_{d,\sigma} \int d^d r \frac{f(\mathbf{x}+\mathbf{r}) - f(\mathbf{x}) }{r^{d + \sigma}},    
\end{equation}
with $\gamma_{d,\sigma} =  \frac{2^{\sigma} \Gamma(\frac{d+ \sigma}{2})}{\pi^{d/2} | \Gamma(- \frac{\sigma}{2})|}$. One can derive an alternative expression for this quantity in the momentum space. In terms of Fourier transform of $f(\mathbf{x})$, $f(\mathbf{q})$, one finds
\begin{equation}
\nabla^{\sigma} f (\mathbf{x}) = - \gamma_{d, \sigma} \int d^d q \ f(\mathbf{q}) \ e^{i \mathbf{q} \cdot \mathbf{x}} \int d^d r \frac{1 - e^{i \mathbf{q} \cdot \mathbf{r}}}{r^{d+ \sigma}}.
\end{equation}
and, exploiting the fact that,
\begin{equation}
\int d^d r \frac{1 - e^{i \mathbf{q} \cdot \mathbf{r}}}{r^{d+ \sigma}} = \gamma_{d,\sigma}^{-1} \ q^{\sigma},
\end{equation}
one have
\begin{equation}
\nabla^{\sigma} f (\mathbf{x}) = - \int d^d q \ q^{\sigma} f(\mathbf{q}) e^{i \mathbf{q} \cdot \mathbf{x}}. 
\end{equation}
(from which, in the limit $\sigma \rightarrow 2$ one recovers the usual behavior of the standard Laplacian). 

In our case, we can notice that the quantity present in Eq\,\eqref{rawSLR}, namely 
\begin{equation}
\int d^2 x \int_{r>a} \frac{d^2 r}{r^{2+ \sigma}} [1 - \cos \left(\theta (\mathbf{x}) - \theta (\mathbf{x} + \mathbf{r} ) \right)] .
\end{equation}
naturally fits into the definition of a two-dimensional fractional Laplacian. Indeed, let us notice that, provided that $\sigma <2$, one can actually extend the integral on $r$ on the whole space, and absorb the contribution coming from the $r<a$ region into the definition of the SR term. Then, we can write the additional LR term as the real part of
\begin{equation}
\begin{split}
\int d^2 x \int \frac{d^2 r}{r^{2+ \sigma}} [1 -  e^{i\theta (\mathbf{x}+ \mathbf{r}) - i\theta (\mathbf{x})}] &= \\
e^{-i\theta (\mathbf{x})} \int d^2 x \int \frac{d^2 r}{r^{2+ \sigma}} [ e^{i\theta (\mathbf{x})} -  e^{i\theta (\mathbf{x}+ \mathbf{r})}] .
\end{split}
\end{equation}
In turn, this can be rewritten by exploiting the definition \eqref{frac} of the fractional Laplacian, 
\begin{equation}
- \gamma_{2,\sigma}^{-1} \int d^2 x \, e^{-i \theta} \nabla^{\sigma} e^{i \theta} .
\end{equation}
The expression above is already real, so that we recover the form of the long-range term given in the main text.

\section{Renormalization procedure} 
\label{AppD}
\noindent
\textcolor{black}{We provide here a derivation of} Eqs.\,\eqref{RG LR} through the RG procedure. Let us consider the action $S[\theta]$ written as 
\begin{equation} \label{StartingAction}
S[\theta] = \int d^2 x \left( \frac{J_{\ell}}{2} | \nabla \theta|^2 + \frac{g_{\ell}}{2} \int_{r>a} \frac{d^2 r}{r^{2+ \sigma}} \left[ 1 - \cos \left( \Delta_{\mathbf{r}} \theta (\mathbf{x}) \right) \right] \right)
\end{equation}
with $\Delta_{\mathbf{r}} \theta(\mathbf{x}) = \theta(\mathbf{x}+\mathbf{r})-\theta(\mathbf{x})$. We are going to compute the RG flow equations perturbatively in $g$ and in the vortex fugacity $y$. Let us notice, however, how the effect of vortices is not encoded in Eq.\,\eqref{StartingAction}. However, at the first perturbative order in $y$ and $g$ the two perturbations act independently, so that we can consider the effect of the renormalization of the vortices on the SR kinetic term only. As usual in this case, one can map this theory into the Sine-Gordon action: 
\begin{equation}
    S_{\rm SG} = \int d^2 \mathbf{x} \left( \frac{1}{2J_{\ell}} |\nabla \varphi|^2 - y_{\ell} \cos 2 \pi \varphi \right) .
\end{equation}
At the first order of the renormalization group, we get that $y$ varies accordingly the scaling dimension of the vertex operator $\cos(2 \pi \varphi)$
\begin{equation} \label{RGy}
    \frac{d y_{\ell}}{d \ell} = (2 - \pi J) y ,
\end{equation}
while $J_{\ell}$ is left unchanged. This is in agreement with the Kosterlitz-Thouless RG for the SR, in which $\dot{J} = O(y^2)$.

Let us now consider the first-order effect of the long-range perturbation: in this case we can set $y=0$. We can then write the field $\theta(\mathbf{x})$ as the sum of a fast-varying and a slow-varying component. In particular, introducing the momentum cutoff $\Lambda = \frac{2 \pi}{a}$, we have $\theta = \theta^{>} + \theta^{<}$ with
\begin{equation}
\begin{split}
\theta^{<} (\mathbf{x}) &= \int_{q < \Lambda e^{- d \ell}} \frac{d^2 q}{(2 \pi)^2} \theta(\mathbf{q}) e^{i \mathbf{q} \cdot \mathbf{x}} \\
\theta^{>} (\mathbf{x}) &= \int_{\Lambda > q > \Lambda e^{- d \ell}} \frac{d^2 q}{(2 \pi)^2} \theta(\mathbf{q}) e^{i \mathbf{q} \cdot \mathbf{x}},
\end{split}
\end{equation}
and integrate out $\theta^{>}$. The non-Gaussian part of the action can be expanded in cumulant: at the first order we have
\begin{equation}
S_{\rm eff} [\theta^{<}] = S_0 [\theta^{<}] +  \me{S_{\rm LR}}_{>}  + O(g^2). 
\end{equation} 
Because of the $\mathbb{Z}_2$ simemtry in $\theta$ symmetry, we can replace $\cos(\Delta_{\mathbf{r}} \theta)$ with $\cos(\Delta_{\mathbf{r}} \theta^{>}) \cos(\Delta_{\mathbf{r}} \theta^{<})$, finding, up to immaterial constants
\begin{equation}
\me{S_{\rm LR}}_{>} =  \frac{g_{\ell}}{2} \int d^2 x  \int \frac{d^2 r}{r^{2+ \sigma}} \me{\cos(\Delta_{\mathbf{r}} \theta^{>})}_{>} \left[ 1 - \cos \left( \Delta_{\mathbf{r}} \theta^{<}  \right) \right]
\end{equation}
(from now on we let the integration run from $r=0$).  Exploiting the identity $\me{\cos(\Delta_{\mathbf{r}} \theta^{>})}_{>} = e^{- \frac{1}{2} \me{\left( \theta(\mathbf{r})- \theta(0) \right)^2}_{>}}$, valid on the Gaussian measure, one has
\begin{equation}
\begin{split}
  \frac{1}{2} &
  \me{\left( \theta(\mathbf{r})- \theta(0) \right)^2}_{>}  \\ &
  = \int_{\Lambda > q > \Lambda e^{- d \ell}} \frac{d^2 q}{(2 \pi)^2} \frac{1 - \cos(\mathbf{q} \cdot \mathbf{r})}{J_{\ell} q^2}  \\ &
  = \frac{d \ell}{2 \pi J_{\ell}} \Big(1 - \mathcal{J}_0(\Lambda r)\Big), 
\end{split}
\end{equation}
where, as before, we denoted with $\mathcal{J}_0 (x)$ the zeroth-order Bessel function of the first kind. Then, remembering that $\eta_{\rm sr} (J) = \frac{1}{2 \pi J_\ell}$, we have 
\begin{equation}
\begin{split}
\me{ \cos(\Delta_{\mathbf{r}} \theta^{>})}_{>}  &= e^{- \eta_{\rm sr}(J_\ell) d \ell \left( 1 - \mathcal{J}_0 (\Lambda r) \right)}  \\ &= 1 -  \eta_{\rm sr}(J_\ell) d \ell + \eta_{\rm sr}(J_\ell) d \ell \mathcal{J}_0(\Lambda r)
\end{split}
\end{equation}
The first two terms provide, as expected, an anomalous dimension of the coupling $g$, namely
$g_{\ell + d \ell} = g_{\ell} e^{- \eta_{\rm sr} (J_\ell) d \ell}$. The last term, however, results in a new term in the action of the form
\begin{equation}
\begin{split}
  \me{S_g}_{>} = & \frac{1}{2} \int d^2 x \Biggl\{ \int \frac{d^2 r}{r^{2+ \sigma}}  g e^{- \eta_{\rm sr}
    (J_\ell) d \ell}  \left[ 1 - \cos \left( \Delta_{\mathbf{r}} \theta^{<} \right) \right] \\
 + &g \eta_{\rm sr}(J_\ell) d \ell   \int \frac{d^2 r}{r^{2+ \sigma}} \mathcal{J}_0(\Lambda r) \left[ 1 - \cos \left( \Delta_{\mathbf{r}} \theta^{<} \right) \right]  \Biggr\}.
\end{split}
\end{equation}
The second term of l.h.s. has the same form of the original $XY$ form. However, for $x>>1$ $\mathcal{J}_0 (x) \sim x^{-1/2} \cos(x - \pi/4)$ so that this effective interaction has
a fast-decaying oscillating behavior. Since this act as a natural cutoff for $r \sim \Lambda^{-1} \sim a$ this can be interpreted as an additional local interaction, and reabsorbed into the SR part of $S[\theta]$. One natural way to procede is to expand $1 - \cos(\Delta_{r} \theta) \approx
\frac{1}{2}(\mathbf{r} \cdot \nabla_{\mathbf{x}} \theta)^2$ so that
\begin{equation}
\int \frac{d^2 r}{r^{2+ \sigma}} \mathcal{J}_0(\Lambda r) (\mathbf{r} \cdot \nabla_{\mathbf{x}} \theta^{<})^2 = \pi |\nabla_{\mathbf{x}} \theta^{<}|^2 \int^{\Lambda^{-1}}_{a} dr r^{1- \sigma} \mathcal{J}_0(\Lambda r) .
\end{equation}
For $\sigma > \frac{1}{2}$ the integral is infrared convergent so that one can neglect the cutoff (this is safe, since we are interested in the $\sigma>7/4$ regime). Finally, putting $r = a u $, we find the correction of the action as 
\begin{equation}
\frac{c_{\sigma}}{2}  (g_{\ell} a^{2- \sigma}) \eta_{\rm sr} (J_\ell) d \ell \int d^2 x |\nabla_{\mathbf{x}} \theta^{<}|^2,
\end{equation}
where
$c_{\sigma} = \frac{\pi}{2} \int_1^{\infty} du
u^{1-\sigma} \mathcal{J}_0(2 \pi u) > 0 $. However, the precise expression of the coefficient is influential important for what follows. 

Summarizing, we found that the effect of the integration over the fast modes can be reabsorbed, at the first order in $g$,$y$, into a redefinition of the couplings $g \rightarrow g + dg$, $J \rightarrow g + dJ$ with
\begin{equation}
\begin{split}
dg &= - \eta_{\rm sr} (J_{\ell}) g_{\ell} d \ell \\
dJ &= c_{\sigma} \eta_{\rm sr} (J_{\ell}) (g_{\ell} a^{2- \sigma})   d \ell. 
\end{split}
\end{equation}
Finally, we ought to perform the rescaling $\mathbf{x} \rightarrow \mathbf{x} e^{-d \ell}$, in order to obtain a theory with the same cutoff as the original one. Then $g$, $J$ are further modified by their own bare length-dimension (namely $2 - \sigma$ and $0$)
\begin{equation}
\begin{split}
dg &= (2 - \sigma - \eta_{\rm sr} (J_{\ell}) ) g_{\ell} d \ell \\
dJ &= c_{\sigma} \eta_{\rm sr} (J_{\ell}) (g_{\ell} a_0^{2- \sigma})   d \ell.
\end{split}
\end{equation}
The RG equations then take the form
\begin{equation} 
\begin{split}
\frac{dg}{d\ell} &=\left(2 - \sigma - \eta_{\rm sr} (J_{\ell}) \right) g_{\ell}  \\
\frac{dJ}{d\ell} &= c_{\sigma} \eta_{\rm sr} (J_{\ell}) g_{\ell}, 
\end{split}
\end{equation}
(we absorbed the dimensional part $a^{2 - \sigma}$ into the definition of the constant $c_{\sigma}$). Together with Eq.\,\eqref{RGy}, those constitute the desired set of equations. 

\section{Derivation of Eq.\,\eqref{gscaling}} 
\label{AppE}
\noindent
Let us consider the flow equations \eqref{RG LR} in the $y=0$ plane. It follows that
\begin{equation}
g_{\ell} = g e^{(2-\sigma) \ell} e^{- \int \eta_{\rm sr} (J_{\ell}) d \ell},  
\end{equation}
which is a reliable result as long as $g_{\ell}$ is small. Now let us consider a RG flow 
\textcolor{black}{which} starts very close to the critical temperature $T_c$: the corresponding trajectory will follow the sepratrix up to the vicinity of the fixed point $g=0$,$J=J_{\sigma}$. According to Eq.\,\eqref{flux}, this is given by the equation
\begin{equation}
g = \frac{\pi (2 - \sigma)}{c_{\sigma}} \left[ (J- J_{\sigma})^2 + k \right]    
\end{equation}
with $k \rightarrow 0^{+}$. We now consider a point in the flow $\ell^{*}$ such that $g(\ell^{*})$ is small and $J(\ell^{*}) > J_{\sigma}$. Then
\begin{equation} \label{integral}
\begin{split}
\int^{\ell^{*}}_0 \eta_{\rm sr} (J_{\ell}) d \ell &= \int^{\ell^{*}}_{J_0} \eta_{\rm sr} (J) \ \frac{dJ}{\dot{J}} \\ 
&= c_{\sigma}^{-1} \int^{\ell^{*}}_{J_0} \frac{dJ}{g(J)} \\
&= \pi(2 - \sigma) \int^{J(\ell^{*})}_{J_0} \frac{dJ}{ \left( J - J_{\sigma}	\right)^2 + k }
\end{split}
\end{equation}
Let us consider now the physical bar\textcolor{black}{e} parameter $J_0$ as a function of the temperature. This will cross the separatrix ($k \rightarrow 0^{+}$) for some $J_c < J_{\sigma}$ (which corresponds to $T_c$) so that $k \sim T_c - T$. In this case the second order singularity $J_{\sigma}$ lies within the integration interval of Eq.\,\eqref{integral}, so that the integral diverges as $k^{-1/2}$ as $k \rightarrow 0^{+}$. Then we have, as expected
\begin{equation}
    g_{\ell^{*}} \sim e^{-B(T-T_c)^{-1/2}}
\end{equation}
where $B$ is a non universal constant. 

\section{Vortex gas representation of the action} 
\label{AppF}
\noindent
\noindent 
We discuss in this Appendix the vortex gas representation of the action \eqref{action}, starting from the decomposition of Eq.\,\eqref{decompos}. At first let us notice how, from the definition of $\theta_0$, $\theta_{\rm top}$ we get a local condition of the form 
\begin{equation}
    \nabla \times \nabla \theta_0 = 0 \hspace{0.5cm} \nabla \times \nabla \theta_{\rm top} = 2 \pi n(\mathbf{x)} 
\end{equation}
where $n(\mathbf{x}) = \sum_k m_k \delta(\mathbf{x} - \mathbf{x}_k)$ is the vortex density. Let us notice that $\nabla \theta_{\rm top}$ is not irrotational because $\theta_{\rm top}$ is not a single variable function. The curl of a vector field, in $d=2$, is a scalar defined as $\nabla \times \mathbf{a} = \epsilon_{ij} \partial_i a_j$, where the Einstein summation convention has been used and $\epsilon_{ij}$ is the rank\textcolor{black}{-}two completely antisymmetric tensor. Exploiting the condition $\nabla \cdot \nabla \theta_{\rm top} = 0$ we can write $\partial_i \theta_{\rm top} = \epsilon_{ij} \partial_j \bar{\theta}$ where $\bar{\theta}$ is singled valued. It follows that 
\begin{equation}
    \nabla^2 \bar{\theta}(\mathbf{x}) = 2 \pi n(\mathbf{x})
\end{equation}
that can be solved introducing the $d=2$ Green function of the Laplacian $G_c (r) = - \frac{1}{2 \pi} \ln r$, finding
\begin{equation} \label{greenfun}
    \bar{\theta}(\mathbf{x}) = 2 \pi \sum_k m_k G_c (|\mathbf{x} - \mathbf{x}_k|)
\end{equation}
In terms of the decomposition \eqref{decompos} then, the kinetic term in action \eqref{action} decouples into the two terms
\begin{equation}
\begin{split}
    \frac{J}{2} |\nabla \theta|^2 &= \frac{J}{2} |\nabla \theta_0|^2 + \frac{J}{2} |\nabla \theta_{\rm top}|^2 \\
    &= \frac{J}{2} |\nabla \theta_0|^2 + \frac{J}{2} |\nabla \bar{\theta}|^2
\end{split}
\end{equation}
Replacing the solution \eqref{greenfun} in the second term we get the usual Coulomb gas interaction, so that $S[\theta]$ can be written as Eq.\,\eqref{top+ntop}.

Let us now integrate out the non-topological field perturbatively in $g$. At the first order in $g$ the cumulant expansion gives the correction: 
\begin{equation}
    \me{S_{\rm LR} [\theta_0 + \theta_{\rm top}]}_0
\end{equation}
where the average is computed on the quadratic non-topological term. By exploiting the form \eqref{rawSLR} of $S_{\rm LR}$, and considering that only the even terms in $\theta_0$ survives we have: 
\begin{equation} \label{intermediate}
\begin{split}
 &\int d^2 \mathbf{x} \int_{r>a} d^2 \mathbf{r} \frac{\me{\cos(\Delta_{\mathbf{r}}\theta_0)}_0}{r^{2+\sigma}} (1-\cos\Delta_{\mathbf{r}} \theta_{\rm top}) \\ &\sim \int d^2 \mathbf{x} \int_{r>a} d^2 \mathbf{r} \frac{1-\cos\Delta_{\mathbf{r}} \theta_{\rm top} (\mathbf{x})}{r^{2+\sigma + \eta_{\rm sr}(J)}}
\end{split}
\end{equation}
up to immaterial additive constants. This has the same form of Eq.\,\eqref{rawSLR} so that, provided that $\sigma + \eta_{\rm sr}(J)<2$, it can be written as well as
\begin{equation}
\begin{split}
    \frac{g}{2} \int d^2 \mathbf{x} \ e^{-i\theta_{\rm top}}  \nabla^{\sigma + \eta_{\rm sr} (J)} e^{i\theta_{\rm top}} 
\end{split}
\end{equation}
where we absorbed some multiplicative proportionality factor into the coupling $g$. If $\sigma + \eta_{\rm sr}(J)>2$, then the dominant term is the Laplacian.

\end{document}